# A unified thermostat scheme for efficient configurational sampling for classical/quantum canonical ensembles *via* molecular dynamics

*Zhijun Zhang*[†], *Xinzijian Liu*[†], *Zifei Chen, Haifeng Zheng, Kangyu Yan, and Jian Liu**

Beijing National Laboratory for Molecular Sciences, Institute of Theoretical and Computational Chemistry, College of Chemistry and Molecular Engineering,

Peking University, Beijing 100871, China

* Electronic mail: jianliupku@pku.edu.cn

† Both authors contributed equally to the work.

## Abstract

We show a unified second-order scheme for constructing simple, robust and accurate algorithms for typical thermostats for configurational sampling for the canonical ensemble. When Langevin dynamics is used, the scheme leads to the BAOAB algorithm that has been recently investigated. We show that the scheme is also useful for other types of thermostat, such as the Andersen thermostat and Nosé-Hoover chain, regardless of whether the thermostat is deterministic or stochastic. In addition to analytical analysis, two 1-dimensional models and three typical realistic molecular systems that range from the gas phase, clusters, to the condensed phase are used in numerical examples for demonstration. Accuracy may be increased by an order of magnitude for estimating coordinate-dependent properties in molecular dynamics (when the same time interval is used), irrespective of which type of thermostat is applied. The scheme is especially useful for path integral molecular dynamics, because it consistently improves the efficiency for evaluating all thermodynamic properties for any type of thermostat.

## I. Introduction

Since the pioneering work of Fermi, Pasta, and Ulam in 1955[1], molecular dynamics (MD) has presented a useful tool for investigating and predicting properties of a wide variety of realistic systems in physics, chemistry, biology, materials, environmental science, *etc*[2, 3]. Various thermostat methods[4-24] have been developed for constant temperature MD simulations. Many cases of them deal with the canonical ensemble where the number of particles (*N*), the volume (*V*), and the temperature (*T*) are constant. Some prevailing thermostats include the Andersen thermostat[5], Langevin dynamics[4, 8, 10, 15, 17, 19-22, 24-26], Nosé-Hoover chain (NHC) [6, 9, 12-14, 27], *etc*. The Andersen thermostat[5] mixes Newtonian dynamics of the particles (of the system) with stochastic collisions with a fictitious heat bath. When a particle is chosen to undergo a collision, its momentum is reselected from the Maxwell-Boltzmann distribution corresponding to the desired temperature *T*. Langevin dynamics offers another type of stochastic thermostat, which is a combination of damping and random perturbation based on Brownian dynamics[4, 8, 10, 15, 17, 19-22, 24-26]. For comparison, the NHC thermostat is deterministic and time-reversible. NHC was developed by Martyna, Tuckerman, and coworkers[12-14, 27] from the original work by Nosé[6] and that by Hoover[9]. It couples the equations of motion of the particles with additional, artificial coordinates and momenta in an extended system approach.

The time interval (time stepsize) controls both the accuracy and efficiency of a MD simulation. While a too small time interval reduces the sampling efficiency in the full phase space, a too large one lowers the accuracy or even breaks down the propagation of the trajectory. The time interval depends on both the system of interest and the integrator/algorithm employed in the MD simulation. It is then both appealing and challenging to develop integrators/algorithms that use larger time intervals to improve the sampling efficiency while maintaining the accuracy.

Most practical thermostat algorithms employ second-order schemes because of its simplicity and efficiency. Higher order schemes that factorize the time interval such as the Suzuki-Yoshida decomposition framework[28-31] may in principle improve the performance but more force calculations are required. In addition, the Suzuki-Yoshida decomposition framework does not perform better than second order schemes when the time interval is large. Other higher-order factorizations such as the Suzuki-Chin factorizations[32-36] require second-order derivatives or even higher order derivatives of the potential energy surface. Unless the potential energy surface or the force field is of some specific forms, it is often much more demanding to obtain its second-order or even higher order derivatives, regardless of whether analytical forms or finite difference techniques are employed (for computing these derivatives). So higher order schemes do not offer more economic algorithms for general molecular systems. In the paper we focus on second-order schemes.

Because all structural properties and most thermodynamic observables only depend on coordinate variables, it is often more important to obtain an accurate sampling in the coordinate space rather than in the momentum space. When MD is used to perform the imaginary time path integral sampling in so called path integral molecular dynamics (PIMD)[37, 38] for quantum canonical ensembles, since all thermodynamic properties depend on only the coordinates of the path integral beads, it is crucial to faithfully sample the configurational distribution (of the path integral beads)[24]. Leimkuhler and Matthews have recently proposed an efficient MD algorithm (‘BAOAB’) for sampling the coordinate space with Langevin dynamics for the canonical ensemble[20, 22]. More recently we have employed BAOAB to develop a simple and accurate algorithm for accomplishing PIMD with the Langevin thermostat[24].

The purpose of this paper is to present a unified scheme that leads to the BAOAB algorithm when Langevin dynamics is used and that may also be applied to other thermostats for efficient configurational sampling for the canonical ensemble. Section II first briefly reviews several typical thermostats such as the Andersen thermostat, Langevin dynamics, and NHC. Section III presents three second-order schemes for the thermostats. Section IV then shows error analysis for the thermostat algorithms in the harmonic limit and that for a general system. Numerical examples are demonstrated in Section V, where thermodynamic properties such as the average potential energy and the average kinetic energy are computed as a function of the time interval[39]. Conclusions and outlook follow in Section VI.

## II. Three typical thermostats for molecular dynamics

Assume the (time-independent) Hamiltonian of the system $H$ to be of standard Cartesian form

$$H = \mathbf{p}^T \mathbf{M}^{-1} \mathbf{p}/2 + U(\mathbf{x}) \quad , \tag{1}$$

where $\mathbf{M}$ is the diagonal 'mass matrix' with elements $\{m_j\}$, and $\mathbf{p}$ and $\mathbf{x}$ are the momentum and coordinate vectors, respectively. $N$ is the number of particles and $3N$ is the total number of degrees of freedom. $T$ is the temperature of the system. ( $\beta = 1/k_B T$ with $k_B$ as the Boltzmann constant.) Below we discuss three typical thermostats often used in MD simulations.

### 1. Langevin dynamics

Langevin dynamics[4, 8, 10, 15, 17, 20-22, 24-26] is a type of a thermostat that employs stochastic dynamics to achieve the desired temperature of the MD simulation. Equations of motion in Langevin dynamics are

$$\begin{pmatrix} \dot{\mathbf{x}} \\ \dot{\mathbf{p}} \end{pmatrix} = \begin{pmatrix} \mathbf{M}^{-1}\mathbf{p} \\ -\dfrac{\partial U(\mathbf{x})}{\partial \mathbf{x}} - \gamma \mathbf{p} + \sqrt{\dfrac{2\gamma}{\beta}}\,\mathbf{M}^{1/2}\tilde{\boldsymbol{\eta}}(t) \end{pmatrix} \quad . \tag{2}$$

Here $\tilde{\boldsymbol{\eta}}(t)$ is a vector. Its element $\tilde{\eta}_j^{(i)}(t)$ is an independent Gaussian-distributed random number with zero mean and unit variance [$\langle \tilde{\eta}_j^{(i)}(t) \rangle = 0$ and $\langle \tilde{\eta}_j^{(i)}(t)\tilde{\eta}_j^{(i)}(t') \rangle = \delta(t-t')$], which is different for each of three degrees of freedom (i.e., *x*, *y*, or *z*) in the 3-dimensional space $(i = 1,2,3)$, each particle $\left(j = \overline{1,N}\right)$, and each time step. The Langevin friction coefficient $\gamma$ is the same for all degrees of freedom $\left(i = \overline{1,3N}\right)$. (Here we consider $\gamma$ as a constant for simplicity. The friction is in general a matrix.)

Eq. (2) is often divided into three parts[20, 22, 24-26, 40]

$$\begin{pmatrix} \dot{\mathbf{x}} \\ \dot{\mathbf{p}} \end{pmatrix} = \underbrace{\begin{pmatrix} \mathbf{M}^{-1}\mathbf{p} \\ 0 \end{pmatrix}} + \underbrace{\begin{pmatrix} 0 \\ -\dfrac{\partial U(\mathbf{x})}{\partial \mathbf{x}} \end{pmatrix}} + \underbrace{\begin{pmatrix} 0 \\ -\gamma \mathbf{p} + \sigma \mathbf{M}^{1/2}\tilde{\boldsymbol{\eta}}(t) \end{pmatrix}} \tag{3}$$

with $\sigma = \sqrt{\dfrac{2\gamma}{\beta}}$ and each of the three parts may be solved 'exactly'. The first part of the right-hand side (RHS) of Eq. (3) for a time interval $\Delta t$ is updating the coordinate

$$\mathbf{x} \leftarrow \mathbf{x} + \mathbf{M}^{-1}\mathbf{p}\Delta t \quad . \tag{4}$$

While the 2nd part of the RHS of Eq. (3) leads to

$$\mathbf{p} \leftarrow \mathbf{p} - \frac{\partial U(\mathbf{x})}{\partial \mathbf{x}} \Delta t \quad , \tag{5}$$

the solution to the third part [i.e., the Ornstein-Uhlenbeck (OU) part] is

$$\mathbf{p} \leftarrow e^{-\gamma \Delta t} \mathbf{p} + \sqrt{\frac{1 - e^{-2\gamma \Delta t}}{\beta}} \mathbf{M}^{1/2} \tilde{\boldsymbol{\eta}} \quad . \tag{6}$$

Here $\tilde{\boldsymbol{\eta}}$ is the independent Gaussian-distributed random number vector as discussed for Eq. (2).

The phase space propagators for the three parts are then $e^{\mathcal{L}_{\mathbf{x}} \Delta t}$, $e^{\mathcal{L}_{\mathbf{p}} \Delta t}$, and $e^{\mathcal{L}_T \Delta t}$, respectively. I.e., the relevant Kolmogorov operators are

$$\mathcal{L}_{\mathbf{x}} \rho = -\mathbf{p}^T \mathbf{M}^{-1} \frac{\partial \rho}{\partial \mathbf{x}} \quad , \tag{7}$$

$$\mathcal{L}_{\mathbf{p}} \rho = \left( \frac{\partial U}{\partial \mathbf{x}} \right)^T \frac{\partial \rho}{\partial \mathbf{p}} \quad , \tag{8}$$

$$\mathcal{L}_T \rho = \frac{\partial}{\partial \mathbf{p}} \cdot (\gamma \mathbf{p} \rho) + \frac{1}{\beta} \frac{\partial}{\partial \mathbf{p}} \cdot \left( \gamma \mathbf{M} \frac{\partial \rho}{\partial \mathbf{p}} \right) \quad . \tag{9}$$

It is trivial to verify that the Boltzmann distribution in the physical phase space

$$\rho_{\text{Boltzmann}}(\mathbf{x}, \mathbf{p}) = \frac{1}{Z_N} \exp\left[ -\beta \left( \mathbf{p}^T \mathbf{M}^{-1} \mathbf{p} / 2 + U(\mathbf{x}) \right) \right] \tag{10}$$

is a stationary state solution to the Fokker-Planck or forward Kolmogorov equation

$$\frac{\partial \rho}{\partial t} = \mathcal{L}\rho = 0 \quad , \tag{11}$$

with the full Kolmogorov operator $\mathcal{L} = \mathcal{L}_{\mathbf{x}} + \mathcal{L}_{\mathbf{p}} + \mathcal{L}_T$. That is, Langevin dynamics is able to sample the canonical ensemble (provided that it is ergodic).

**2. Andersen thermostat**

The Andersen thermostat[5] is a type of a thermostat that uses stochastic coupling to impose the desired temperature in the MD simulation. In the Andersen thermostat, each particle of the system stochastically collides with a fictitious heat bath, and once the collision occurs, the momentum of this particle is chosen afresh from the Maxwell-Boltzmann momentum distribution. Times between collisions with the heat bath are selected from a Poisson distribution, i.e., the probability distribution is $P(t;\nu) = \nu e^{-\nu t}$, where the collision frequency $\nu$ specifies the coupling strength between the particle and the heat bath. Between stochastic collisions, the propagation of the MD trajectory is at constant energy according to the Hamilton equations of motion or the Newtonian laws of motion. Below we revisit the Andersen thermostat.

The collision step in the algorithm is often described as

Randomly select a number of particles to undergo a collision with the heat bath. The probability that a particle is selected in the time interval $\Delta t$ is $\nu\Delta t$ (more accurately, $1 - e^{-\nu\Delta t}$). If particle *j* is selected, its new momentum is reselected from a Maxwell momentum distribution at the desired temperature *T*, while all other particles are unaffected by this collision.

Note that the explicit form for the collision step at a time interval $\Delta t$ can be expressed as

$$\mathbf{p}^{(j)} \leftarrow \sqrt{\frac{1}{\beta}}\mathbf{M}_j^{1/2}\boldsymbol{\theta}_j, \quad \text{if } \mu_j < \nu\Delta t \left(\text{or more precisely } \mu_j < 1-e^{-\nu\Delta t}\right) \quad \left(j=\overline{1,N}\right) . \tag{12}$$

Here $\mathbf{p}^{(j)}$ is the 3-dimensional momentum vector and $\mathbf{M}_j$ the $3\times3$ diagonal mass matrix for particle $j$. $\mu_j$ is a uniformly distributed random number in the range (0,1), which is different for each particle $\left(j=\overline{1,N}\right)$, and each time when Eq. (12) is invoked. $\boldsymbol{\theta}_j$ is a 3-dimensional vector. Its element $\theta_j^{(i)}(t)$ is an independent Gaussian-distributed random number with zero mean and unit variance, which is different for each of three degrees of freedom (i.e., *x*, *y*, or *z*) in the 3-dimensional space $(i=1,2,3)$, each particle $\left(j=\overline{1,N}\right)$, and each invocation of Eq. (12).

Use $e^{\mathcal{L}_T\Delta t}$ to represent the phase space propagator for the thermostat step at a time interval $\Delta t$. Propagation of the density distribution in the phase space $\rho \equiv \rho(\mathbf{x},\mathbf{p})$ for the collision process can be characterized by the forward Kolmogorov equation

$$\frac{\partial \rho}{\partial t} = \mathcal{L}_T \rho = \nu\left[\rho_{\text{MB}}(\mathbf{p})\int_{-\infty}^{\infty}\rho(\mathbf{x},\mathbf{p})d\mathbf{p} - \rho(\mathbf{x},\mathbf{p})\right] . \tag{13}$$

Here $\rho_{\text{MB}}(\mathbf{p})$ is the Maxwell (or Maxwell-Boltzmann) momentum distribution

$$\rho_{\text{MB}}(\mathbf{p}) = \left(\frac{\beta}{2\pi}\right)^{3N/2}|\mathbf{M}|^{-1/2}\exp\left[-\frac{\beta}{2}\mathbf{p}^T\mathbf{M}^{-1}\mathbf{p}\right] . \tag{14}$$

Using Eqs. (7), (8), and (13), one finds that the full Kolmogorov operator $\mathcal{L} = \mathcal{L}_{\mathbf{x}} + \mathcal{L}_{\mathbf{p}} + \mathcal{L}_T$ for the Andersen thermostat satisfies

$$\mathcal{L}\rho = \nu\left[\rho_{\mathrm{MB}}\left(\mathbf{p}\right)\int_{-\infty}^{\infty}\rho\left(\mathbf{x},\mathbf{p}\right)d\mathbf{p} - \rho\left(\mathbf{x},\mathbf{p}\right)\right] - \mathbf{p}^{T}\mathbf{M}^{-1}\frac{\partial\rho}{\partial\mathbf{x}} + \left(\frac{\partial U}{\partial\mathbf{x}}\right)^{T}\frac{\partial\rho}{\partial\mathbf{p}} \quad . \tag{15}$$

It is straightforward to show that the Boltzmann distribution in the physical phase space [Eq. (10)] is a stationary state solution to the Fokker-Planck or forward Kolmogorov equation Eq. (11) with the full Kolmogorov operator given by Eq. (15). I.e., the Andersen thermostat is able to generate the canonical ensemble (if ergodicity is guaranteed), a well-known statement from Refs. [3, 5, 41].

Integration over time in Eq. (13) leads to

$$e^{\mathcal{L}_T\Delta t}\rho = \left(1-e^{-\nu\Delta t}\right)\rho_{\mathrm{MB}}\left(\mathbf{p}\right)\int_{-\infty}^{\infty}\rho\left(\mathbf{x},\mathbf{p}\right)d\mathbf{p} + e^{-\nu\Delta t}\rho\left(\mathbf{x},\mathbf{p}\right) \quad . \tag{16}$$

It is much more convenient to use Eq. (13) or Eq. (16) to present the analytical analysis for the Andersen thermostat. Note that when $\nu\Delta t$ is small, an approximation of Eq. (16) produces

$$e^{\mathcal{L}_T\Delta t}\rho = \nu\Delta t\rho_{\mathrm{MB}}\left(\mathbf{p}\right)\int_{-\infty}^{\infty}\rho\left(\mathbf{x},\mathbf{p}\right)d\mathbf{p} + \left(1-\nu\Delta t\right)\rho\left(\mathbf{x},\mathbf{p}\right) \quad , \tag{17}$$

which corresponds to the conventional description for the collision step in the Andersen thermostat[3, 5].

### 3. Nosé-Hoover chain

Nosé-Hoover chain (NHC) [6, 9, 12-14, 27] is a type of a thermostat that performs deterministic MD in an extended-system approach to control the temperature in the simulation.

The equations of motion of NHC[27] read

$$\left.\begin{aligned}
\dot{x}_i &= \frac{p_i}{m_i} \\
\dot{p}_i &= -\frac{\partial U}{\partial x_i} - \frac{p_{\eta_1^{(i)}}}{Q_1} p_i \\
\dot{\eta}_j^{(i)} &= \frac{p_{\eta_j^{(i)}}}{Q_j} \\
\dot{p}_{\eta_1^{(i)}} &= \frac{p_i^2}{m_i} - k_B T - \frac{p_{\eta_2^{(i)}}}{Q_2} p_{\eta_1^{(i)}} \\
\dot{p}_{\eta_j^{(i)}} &= \frac{p_{\eta_{j-1}^{(i)}}^2}{Q_{j-1}} - k_B T - \frac{p_{\eta_{j+1}^{(i)}}}{Q_{j+1}} p_{\eta_j^{(i)}} \quad \left(j = \overline{2, M_{\mathrm{NHC}} - 1}\right) \\
\dot{p}_{\eta_{M_{\mathrm{NHC}}}^{(i)}} &= \frac{p_{\eta_{M_{\mathrm{NHC}}-1}^{(i)}}^2}{Q_{M_{\mathrm{NHC}}-1}} - k_B T
\end{aligned}\right\} \left(i = \overline{1, 3N}\right), \tag{18}$$

where $M_{\mathrm{NHC}}$ pairs of additional variables $\left\{\eta_j^{(i)}, p_{\eta_j^{(i)}}\right\}$ $\left(j = \overline{1, M_{\mathrm{NHC}}}\right)$ in a so-called 'Nosé-Hoover chain' are coupled to each physical degree of freedom $\left(i = \overline{1, 3N}\right)$, the parameters $Q_1, \cdots, Q_{M_{\mathrm{NHC}}}$ are called the NHC thermostat masses[12, 14]. An optimal choice for the NHC thermostat masses suggested by Martyna, Tuckerman, and coworkers[12, 14] is

$$Q_j = k_B T \tilde{\tau}_{\mathrm{NHC}}^2 \quad \left(j = \overline{1, M_{\mathrm{NHC}}}\right) \ , \tag{19}$$

where $\tilde{\tau}_{\mathrm{NHC}}$ is the characteristic time of the system. It is claimed in Ref. [12] that the choice of $\tilde{\tau}_{\mathrm{NHC}}$ in NHC is much less critical than that in the Nosé–Hoover method (i.e., $M_{\mathrm{NHC}} = 1$ in NHC).

For the equations of motion in Eq. (18), the conserved quantity is

$$H' = \frac{1}{2}\mathbf{p}^T \mathbf{M}^{-1} \mathbf{p} + U(\mathbf{x}) + \sum_{i=1}^{3N} \sum_{j=1}^{M_{\mathrm{NHC}}} \left( \frac{p_{\eta_j^{(i)}}^2}{2Q_j} + k_B T \eta_j^{(i)} \right) . \tag{20}$$

Eq. (20) is the Hamiltonian for an extended system. I.e., the real system is extended by addition of artificial degrees of freedom. Note that Eq. (18) can not be derived from the Hamilton equations of motion from Eq. (20). Instead, Eq. (18) is a kind of non-Hamiltonian dynamics, in which the phase space volume of the extend-system is not preserved during the propagation. The evolution of the phase space volume satisfies

$$\begin{aligned} d\mathbf{x}_t d\mathbf{p}_t d\boldsymbol{\eta}_t d\mathbf{p}_{\boldsymbol{\eta},t} &= \left| \frac{\partial\left(\mathbf{x}_t, \mathbf{p}_t, \boldsymbol{\eta}_t, \mathbf{p}_{\boldsymbol{\eta},t}\right)}{\partial\left(\mathbf{x}_0, \mathbf{p}_0, \boldsymbol{\eta}_0, \mathbf{p}_{\boldsymbol{\eta},0}\right)} \right| d\mathbf{x}_0 d\mathbf{p}_0 d\boldsymbol{\eta}_0 d\mathbf{p}_{\boldsymbol{\eta},0} \\ &= \exp\left[ \int_0^t \frac{\partial}{\partial \mathbf{x}} \cdot \dot{\mathbf{x}} + \frac{\partial}{\partial \mathbf{p}} \cdot \dot{\mathbf{p}} + \sum_{i=1}^{3N} \sum_{j=1}^{M_{\mathrm{NHC}}} \left( \frac{\partial \dot{\eta}_j^{(i)}}{\partial \eta_j^{(i)}} + \frac{\partial \dot{p}_{\eta_j^{(i)}}}{\partial p_{\eta_j^{(i)}}} \right) dt \right] d\mathbf{x}_0 d\mathbf{p}_0 d\boldsymbol{\eta}_0 d\mathbf{p}_{\boldsymbol{\eta},0} \end{aligned} \quad . \tag{21}$$

Here $\boldsymbol{\eta} \equiv \left\{ \eta_j^{(i)} \middle| j = \overline{1, M_{\mathrm{NHC}}}, i = \overline{1, 3N} \right\}$ and $\mathbf{p}_{\boldsymbol{\eta}} \equiv \left\{ p_{\eta_j^{(i)}} \middle| j = \overline{1, M_{\mathrm{NHC}}}, i = \overline{1, 3N} \right\}$. Substituting Eq. (18) into Eq. (21) leads to

$$\left| \frac{\partial\left(\mathbf{x}_t, \mathbf{p}_t, \boldsymbol{\eta}_t, \mathbf{p}_{\boldsymbol{\eta},t}\right)}{\partial\left(\mathbf{x}_0, \mathbf{p}_0, \boldsymbol{\eta}_0, \mathbf{p}_{\boldsymbol{\eta},0}\right)} \right| = \exp\left[ -\sum_{i=1}^{3N} \sum_{j=1}^{M_{\mathrm{NHC}}} \eta_j^{(i)}(t) + \sum_{i=1}^{3N} \sum_{j=1}^{M_{\mathrm{NHC}}} \eta_j^{(i)}(0) \right] , \tag{22}$$

or equivalently,

$$\exp\left[ \sum_{i=1}^{3N} \sum_{j=1}^{M_{\mathrm{NHC}}} \eta_j^{(i)}(t) \right] d\mathbf{x}_t d\mathbf{p}_t d\boldsymbol{\eta}_t d\mathbf{p}_{\boldsymbol{\eta},t} = \exp\left[ \sum_{i=1}^{3N} \sum_{j=1}^{M_{\mathrm{NHC}}} \eta_j^{(i)}(0) \right] d\mathbf{x}_0 d\mathbf{p}_0 d\boldsymbol{\eta}_0 d\mathbf{p}_{\boldsymbol{\eta},0} \ . \tag{23}$$

That is, the weighted phase space volume is conserved. The microcanonical partition function can then be constructed by using the weighted phase space volume [Eq. (23)] and the conserved quantity [Eq. (20)], which produces

$$Z=\int d\mathbf{x}d\mathbf{p}d\boldsymbol{\eta}d\mathbf{p}_{\boldsymbol{\eta}}\exp\left(\sum_{i=1}^{3N}\sum_{j=1}^{M_{\mathrm{NHC}}}\eta_j^{(i)}\right)\delta\left(\frac{1}{2}\mathbf{p}^T\mathbf{M}^{-1}\mathbf{p}+U\left(\mathbf{x}\right)+\sum_{i=1}^{3N}\sum_{j=1}^{M_{\mathrm{NHC}}}\left(\frac{p_{\eta_j^{(i)}}^2}{2Q_j}+k_BT\eta_j^{(i)}\right)-C\right) , \tag{24}$$

with $C$ a constant. Integration over $\boldsymbol{\eta}$ in Eq. (24) reaches

$$Z=\frac{e^{\beta C}}{k_BTV_\eta^{3NM_{\mathrm{NHC}}-1}}\int d\mathbf{x}d\mathbf{p}d\mathbf{p}_{\boldsymbol{\eta}}\exp\left[-\beta\left(\frac{1}{2}\mathbf{p}^T\mathbf{M}^{-1}\mathbf{p}+U\left(\mathbf{x}\right)+\sum_{i=1}^{3N}\sum_{j=1}^{M_{\mathrm{NHC}}}\frac{p_{\eta_j^{(i)}}^2}{2Q_j}\right)\right] , \tag{25}$$

where $V_\eta$ represents the volume of the 1-dimensional space for each $\eta_j^{(i)}\left(j=\overline{1,M_{\mathrm{NHC}}},i=\overline{1,3N}\right)$.

Integration over $\mathbf{p}_{\boldsymbol{\eta}}$ further leads to

$$Z=\frac{e^{\beta C}}{k_BTV_\eta^{3NM_{\mathrm{NHC}}-1}}\left(\prod_{j=1}^{M_{\mathrm{NHC}}}\frac{2\pi Q_j}{\beta}\right)^{3N/2}\int d\mathbf{x}d\mathbf{p}\exp\left[-\beta\left(\frac{1}{2}\mathbf{p}^T\mathbf{M}^{-1}\mathbf{p}+U\left(\mathbf{x}\right)\right)\right] , \tag{26}$$

which is the product of the canonical partition function (of the physical phase space) and a constant factor. That is, the NHC thermostat in principle produces the exact canonical distribution for the system (provided that it is ergodic), as shown in Ref. [42]. Note that the auxiliary variables $\boldsymbol{\eta}$ are redundant for the dynamics in NHC [i.e., Eq. (18)]. They are used in the equations of motion only for monitoring the conserved quantity Eq. (20). The framework in Eq. (18) is known as the 'massive' thermostat[27, 43], which is employed throughout this paper. Similarly, one can couple a Nosé-Hoover chain to each particle (the 'local' thermostat), or couple it to the whole system (the 'global' thermostat)[27].

## III. Three typical thermostat schemes

Numerical MD integrators for a time interval $\Delta t$ are often consisted of a step for updating the coordinate $\mathbf{x}(t+\Delta t)\leftarrow\mathbf{x}(t)+\mathbf{M}^{-1}\mathbf{p}(t)\Delta t$ , that for updating the momentum $\mathbf{p}(t+\Delta t) \leftarrow \mathbf{p}(t) - U'(\mathbf{x}(t))\Delta t$, and that for the thermostat process that controls the temperature. Use $e^{\mathcal{L}_{\mathbf{x}}\Delta t}$, $e^{\mathcal{L}_{\mathbf{p}}\Delta t}$, and $e^{\mathcal{L}_{T}\Delta t}$ to represent the phase space propagators for the three steps, respectively. Here $\mathcal{L}_{\mathbf{x}}$, $\mathcal{L}_{\mathbf{p}}$, and $\mathcal{L}_{T}$ are the relevant Kolmogorov operators. For instance, $\mathcal{L}_{\mathbf{x}}\rho = -\mathbf{p}^T\mathbf{M}^{-1}\dfrac{\partial \rho}{\partial \mathbf{x}}$ and $\mathcal{L}_{\mathbf{p}}\rho = \left(\dfrac{\partial U}{\partial \mathbf{x}}\right)^T \dfrac{\partial \rho}{\partial \mathbf{p}}$, where $\rho$ is a density distribution in the phase space.

Efficient thermostat MD integrators for a time interval $\Delta t$ were often suggested to be of the form

$$e^{\mathcal{L}\Delta t} \approx e^{\mathcal{L}^{\mathrm{Side}}\Delta t} = e^{\mathcal{L}_T \Delta t/2} e^{\mathcal{L}_{\mathbf{p}} \Delta t/2} e^{\mathcal{L}_{\mathbf{x}} \Delta t} e^{\mathcal{L}_{\mathbf{p}} \Delta t/2} e^{\mathcal{L}_T \Delta t/2} \quad . \tag{27}$$

I.e., the thermostat step is applied for half an interval $\Delta t/2$ before and after a whole step of the velocity Verlet algorithm for constant energy MD is implemented. As the thermostat process is arranged at both the beginning and end of each time interval, we note it the 'side' scheme. Such as the NHC algorithm proposed by Martyna, Tuckerman, and coworkers[12, 14, 27, 44] and the Langevin dynamics algorithm proposed by Bussi *et al.*[25] fall into the category. The path integral Langevin equation (PILE) thermostat recently developed by Ceriotti *et al.*[26] also employed Langevin dynamics in the 'side' scheme for sampling the canonical distribution for PIMD. The numerical examples presented by Ceriotti *et al.* [26] demonstrate that in terms of sampling efficiency PILE is comparable to the NHC algorithm of Tuckerman *et al.* [27, 45, 46] for PIMD. That is, Langevin dynamics is comparable to NHC for sampling the quantum canonical ensemble *via* PIMD when the 'side' scheme is employed. For convenience, when the 'side' scheme is employed in the Andersen thermostat, Langevin dynamics, and NHC, we denote the algorithms 'side-Andersen', 'side-Langevin', and 'side-NHC', respectively.

Close to the 'side' scheme, another scheme was used even earlier

$$e^{\mathcal{L}\Delta t} \approx e^{\mathcal{L}^{\text{End}}\Delta t} = e^{\mathcal{L}_T\Delta t} e^{\mathcal{L}_{\mathbf{p}}\Delta t/2} e^{\mathcal{L}_{\mathbf{x}}\Delta t} e^{\mathcal{L}_{\mathbf{p}}\Delta t/2} \quad . \tag{28}$$

I.e., the thermostat process is applied after a whole step of the velocity Verlet algorithm is implemented. As the thermostat procedure is only used at the end of each time interval, we note it the 'end' scheme. E.g., the original algorithm for the Andersen thermostat[3, 5] in 1980 employed the 'end' scheme. When the 'end' scheme is used in the Andersen thermostat, Langevin dynamics, and NHC, we denote the algorithms 'end-Andersen', 'end-Langevin', and 'end-NHC', respectively.

When the thermostat MD integrators are of the form

$$e^{\mathcal{L}\Delta t} \approx e^{\mathcal{L}^{\text{Middle}}\Delta t} = e^{\mathcal{L}_{\mathbf{p}}\Delta t/2} e^{\mathcal{L}_{\mathbf{x}}\Delta t/2} e^{\mathcal{L}_T\Delta t} e^{\mathcal{L}_{\mathbf{x}}\Delta t/2} e^{\mathcal{L}_{\mathbf{p}}\Delta t/2} \quad , \tag{29}$$

i.e., the thermostat is arranged in the middle, we note it the 'middle' scheme. It also leads to the velocity-Verlet algorithm for constant-energy MD when the thermostat vanishes. The 'middle' scheme [Eq. (29)] has already been proposed for the Langevin thermostat for MD[20] and for PIMD[24]. It has already been shown that Langevin dynamics with Eq. (29) greatly improve the efficiency in sampling the coordinate space in MD[22] and in sampling the configurational distribution of path integral beads in PIMD[24]. It is *important* to note that the 'middle' scheme may be generalized to other thermostats for either MD or PIMD. When the 'middle' scheme is applied in the Andersen thermostat, Langevin dynamics, and NHC, we denote the algorithms 'middle-Andersen', 'middle-Langevin', and 'middle-NHC', respectively.

The thermostat algorithms in the three typical schemes are described in detail in Appendix C.

## IV. Error analysis for different thermostat algorithms

## 1. Stationary state distribution for a harmonic system for a finite time interval

Consider a harmonic system where the potential energy function is

$$U(\mathbf{x}) = \left(\mathbf{x} - \mathbf{x}_{eq}\right)^T \mathbf{A}\left(\mathbf{x} - \mathbf{x}_{eq}\right)/2 \quad . \tag{30}$$

Eq. (8) then becomes

$$\mathcal{L}_{\mathbf{p}} \rho = (\mathbf{x} - \mathbf{x}_{eq})^T \mathbf{A} \frac{\partial \rho}{\partial \mathbf{p}} \quad . \tag{31}$$

Eq. (7) and the Taylor expansion $e^{\mathcal{L}_{\mathbf{x}} \Delta t} = \sum_{n=0}^{\infty} \frac{1}{n!} \left( -\mathbf{p}^T \mathbf{M}^{-1} \Delta t \frac{\partial}{\partial \mathbf{x}} \right)^n$ lead to a shift operator

$$e^{\mathcal{L}_{\mathbf{x}} \Delta t} f(\mathbf{x}) = f(\mathbf{x} - \mathbf{M}^{-1} \mathbf{p} \Delta t) \quad . \tag{32}$$

Similarly, one obtains

$$e^{\mathcal{L}_{\mathbf{p}} \Delta t} g(\mathbf{p}) = g\left(\mathbf{p} + \mathbf{A}\left(\mathbf{x} - \mathbf{x}_{eq}\right) \Delta t\right) \quad . \tag{33}$$

a) Andersen thermostat

Appendix A presents the derivation of the stationary state distribution for a 1-dimensional harmonic system for a finite time interval $\Delta t$. Below we show the multi-dimensional case.

Here we adopt the strategy proposed in Appendix C of Ref. [24]. When the Andersen thermostat is used, the collision process [Eq. (12) or Eq. (16)] leaves the Maxwell momentum distribution unchanged, i.e.,

$$e^{\mathcal{L}_T \Delta t} \exp\left\{ -\beta \left[ \frac{1}{2} \mathbf{p}^T \mathbf{M}^{-1} \mathbf{p} \right] \right\} = \exp\left\{ -\beta \left[ \frac{1}{2} \mathbf{p}^T \mathbf{M}^{-1} \mathbf{p} \right] \right\} \quad . \tag{34}$$

Consider the density distribution

$$\rho^{\text{Side}} = \frac{1}{Z_N} \exp\left[ -\beta \left( \frac{1}{2}\mathbf{p}^T\mathbf{M}^{-1}\mathbf{p} + \frac{1}{2}\left(\mathbf{x}-\mathbf{x}_{eq}\right)^T (\mathbf{1}-\mathbf{A}\mathbf{M}^{-1}\frac{\Delta t^2}{4})\mathbf{A}\left(\mathbf{x}-\mathbf{x}_{eq}\right) \right) \right] , \tag{35}$$

where $Z_N$ is the normalization constant. Using Eq. (27) and Eqs. (31)-(34), it is straightforward to verify

$$e^{\mathcal{L}^{\text{Side}}\Delta t} \rho^{\text{Side}} = \rho^{\text{Side}} \quad . \tag{36}$$

I.e., Eq. (35) is the stationary state distribution for the 'side' scheme.

Similarly, while the stationary state distribution for the 'end' scheme for the harmonic system is the same as Eq. (35), i.e.,

$$\rho^{\text{End}} = \frac{1}{Z_N} \exp\left[ -\beta \left( \frac{1}{2}\mathbf{p}^T\mathbf{M}^{-1}\mathbf{p} + \frac{1}{2}\left(\mathbf{x}-\mathbf{x}_{eq}\right)^T (\mathbf{1}-\mathbf{A}\mathbf{M}^{-1}\frac{\Delta t^2}{4})\mathbf{A}\left(\mathbf{x}-\mathbf{x}_{eq}\right) \right) \right] , \tag{37}$$

that for the 'middle' scheme is

$$\rho^{\text{Middle}} = \frac{1}{\bar{Z}_N} \exp\left[ -\beta \left( \frac{1}{2}\mathbf{p}^T (\mathbf{M}-\mathbf{A}\frac{\Delta t^2}{4})^{-1}\mathbf{p} + \frac{1}{2}\left(\mathbf{x}-\mathbf{x}_{eq}\right)^T \mathbf{A}\left(\mathbf{x}-\mathbf{x}_{eq}\right) \right) \right] , \tag{38}$$

where $\bar{Z}_N$ is the normalization constant.

When the time interval $\Delta t$ is finite, both the 'side' and 'end' schemes produce the exact momentum distribution but not the exact configurational distribution in harmonic limit. For comparison, the ‘middle’ scheme leads to the exact configurational distribution but not the exact momentum distribution for the harmonic system.

b) Langevin dynamics

When Langevin dynamics is employed as the thermostat, the OU process [Eq. (6) or Eq. (9)] keeps the Maxwell momentum distribution unchanged[24]. That is, Eq. (34) also holds in Langevin dynamics[24]. It is then trivial to show that the Andersen thermostat and Langevin dynamics approach the same stationary state distribution in the harmonic limit, when the same scheme is applied. The conclusion holds for any other thermostats as long as they also keep the Maxwell momentum distribution unchanged in the thermostat step.

c) Nosé-Hoover chain

We first consider that the exact phase space propagator $e^{\mathcal{L}_T \Delta t}$ for the NHC thermostat part [Eq. (138)] were available. It is then straightforward to verify the propagator for the thermostat part [Eq. (138)] keeps the Maxwell-Boltzmann distribution for both the physical momentum and the auxiliary momentum variables unchanged, i.e.,

$$e^{\mathcal{L}_T \Delta t} \exp\left\{-\beta\left[\frac{1}{2}\mathbf{p}^T\mathbf{M}^{-1}\mathbf{p} + \sum_{i=1}^{3N}\sum_{j=1}^{M_{\mathrm{NHC}}} \frac{p_{\eta_j^{(i)}}^2}{2Q_j}\right]\right\} = \exp\left\{-\beta\left[\frac{1}{2}\mathbf{p}^T\mathbf{M}^{-1}\mathbf{p} + \sum_{i=1}^{3N}\sum_{j=1}^{M_{\mathrm{NHC}}} \frac{p_{\eta_j^{(i)}}^2}{2Q_j}\right]\right\} . \quad (39)$$

The stationary state marginal distribution of the variables $\left(\mathbf{x}, \mathbf{p}, \mathbf{p}_{\boldsymbol{\eta}}\right)$ for the harmonic system obtained by ‘side-NHC’ is

$$\rho^{\text{Side-NHC}} = \frac{1}{Z_N{}'} \exp\left[-\beta\left(\frac{1}{2}\mathbf{p}^T\mathbf{M}^{-1}\mathbf{p} + \frac{1}{2}\left(\mathbf{x}-\mathbf{x}_{eq}\right)^T (\mathbf{1} - \mathbf{A}\mathbf{M}^{-1}\frac{\Delta t^2}{4})\mathbf{A}\left(\mathbf{x}-\mathbf{x}_{eq}\right) + \sum_{i=1}^{3N}\sum_{j=1}^{M_{\mathrm{NHC}}} \frac{p_{\eta_j^{(i)}}^2}{2Q_j}\right)\right] . \quad (40)$$

Here, $Z_N'$ is the normalization constant. Integration over $\mathbf{p}_{\boldsymbol{\eta}}$ in Eq. (40) leads to the stationary state marginal distribution for the physical phase space variables $(\mathbf{x},\mathbf{p})$, which is the same as Eq. (35). Similarly, the stationary state marginal distribution of the physical phase space variables $(\mathbf{x},\mathbf{p})$ for the harmonic system obtained by 'end-NHC' also leads to Eq. (35), while that produced by 'middle-NHC' is the same as Eq. (38).

Although the analytical solution for the exact phase space propagator $e^{\mathcal{L}_T \Delta t}$ for the NHC thermostat part [Eq. (138)] is difficult to obtain, the multiple time-scale scheme such as the reference system propagator algorithm[13] (RESPA) and a higher-order (than $\Delta t^2$) factorization such as the Suzuki-Yoshida decomposition framework[28-30] and the optimized Forest–Ruth-like algorithm[47] can be applied to the NHC thermostat part to achieve effectively accurate numerical results. Note that the higher-order (than $\Delta t^2$) factorization is only used for the NHC thermostat part, not for the physical degrees of freedom. The numerical performance of NHC is in practice similar to that of Langevin dynamics or the Andersen thermostat.

### 2. Comparison between the 'side' and 'end' schemes for a general system

We compare the accuracy of the 'side' scheme [Eq. (27)] to that of the 'end' scheme [Eq. (28)] for a general system.

1) Andersen thermostat

We first consider the Andersen thermostat. Note that Eq. (16) is an exact solution to Eq. (13), the Fokker-Planck or forward Kolmogorov equation for the collision process in the Andersen thermostat. We first prove the equality

$$e^{\mathcal{L}_T \Delta t} \rho(\mathbf{p}_0;0) = e^{\mathcal{L}_T \Delta t/2} e^{\mathcal{L}_T \Delta t/2} \rho(\mathbf{p}_0;0) \quad . \tag{41}$$

Here $\rho(\mathbf{p}_0;0)$ is an arbitrary probability distribution of $\mathbf{p}_0$ at time 0. The Kolmogorov operator $\mathcal{L}_T$ for the collision process is defined in Eq. (13). The left-hand side (LHS) of Eq. (41) can be expressed as

$$e^{\mathcal{L}_T \Delta t} \rho(\mathbf{p}_0;0) = \int d\mathbf{p}_0 \rho(\mathbf{p};\Delta t|\mathbf{p}_0;0) \rho(\mathbf{p}_0;0) \quad . \tag{42}$$

Here $\rho(\mathbf{p};\Delta t|\mathbf{p}_0;0)$ is the conditional probability distribution of $\mathbf{p}$ at time $\Delta t$ given $\mathbf{p}_0$ at time 0. Eq. (13) leads to

$$\rho(\mathbf{p};\Delta t|\mathbf{p}_0;0) = e^{-\nu\Delta t} \delta(\mathbf{p}-\mathbf{p}_0) + \left(1-e^{-\nu\Delta t}\right) \rho_{MB}(\mathbf{p}) \ , \tag{43}$$

an exact solution for the Fokker-Planck or forward Kolmogorov equation

$$\frac{\partial}{\partial t} \rho = \mathcal{L}_T \rho \tag{44}$$

for the collision process in the Andersen thermostat. It is then trivial to show that

$$\rho(\mathbf{p};\Delta t|\mathbf{p}_0;0) = \int \rho(\mathbf{p};\Delta t|\mathbf{p}_1;\Delta t/2) \rho(\mathbf{p}_1;\Delta t/2|\mathbf{p}_0;0) d\mathbf{p}_1 \quad , \tag{45}$$

which produces Eq. (41). It is then straightforward to verify that the stationary state distribution of 'side-Andersen' and that of 'end-Andersen' have the relation

$$\rho^{\text{End-ADS}}(\mathbf{x},\mathbf{p}) = e^{\mathcal{L}_T \Delta t/2} \rho^{\text{Side-ADS}}(\mathbf{x},\mathbf{p}) \ . \tag{46}$$

Because the Andersen thermostat does *not* change the marginal distribution of $\mathbf{x}$, 'end-Andersen' and 'side-Andersen' share the same stationary state marginal distribution of the coordinate

$$\rho_{\mathbf{x}}^{\text{End-ADS}}(\mathbf{x}) = \rho_{\mathbf{x}}^{\text{Side-ADS}}(\mathbf{x}) \ . \tag{47}$$

Integration over $\mathbf{x}$ in Eq. (46) produces

$$\rho_{\mathbf{p}}^{\text{End-ADS}}\left(\mathbf{p}\right)=e^{\mathcal{L}_T\Delta t/2}\rho_{\mathbf{p}}^{\text{Side-ADS}}\left(\mathbf{p}\right) \ . \tag{48}$$

Implementing Eq. (16), one obtains

$$\rho_{\mathbf{p}}^{\text{End-ADS}}\left(\mathbf{p}\right)=e^{-\nu\Delta t/2}\rho_{\mathbf{p}}^{\text{Side-ADS}}\left(\mathbf{p}\right)+\left(1-e^{-\nu\Delta t/2}\right)\rho_{\text{MB}}\left(\mathbf{p}\right) \tag{49}$$

from Eq. (48). Rearranging Eq. (49) leads to

$$\rho_{\mathbf{p}}^{\text{End-ADS}}\left(\mathbf{p}\right)-\rho_{\text{MB}}\left(\mathbf{p}\right)=e^{-\nu\Delta t/2}\left[\rho_{\mathbf{p}}^{\text{Side-ADS}}\left(\mathbf{p}\right)-\rho_{\text{MB}}\left(\mathbf{p}\right)\right] \ . \tag{50}$$

Taking the absolute value in Eq. (50), one finds

$$\left|\rho_{\mathbf{p}}^{\text{End-ADS}}\left(\mathbf{p}\right)-\rho_{\text{MB}}\left(\mathbf{p}\right)\right|=e^{-\nu\Delta t/2}\left|\rho_{\mathbf{p}}^{\text{Side-ADS}}\left(\mathbf{p}\right)-\rho_{\text{MB}}\left(\mathbf{p}\right)\right| \ . \tag{51}$$

Since the inequality $e^{-\nu\Delta t/2}\le 1$ always holds, the stationary state marginal distribution of the momentum produced by ‘end-Andersen’ is not less accurate than that obtained by ‘side-Andersen’.

Consider the averaged kinetic energy produced by ‘end-Andersen’

$$\frac{1}{2}\left\langle\mathbf{p}^T\mathbf{M}^{-1}\mathbf{p}\right\rangle_{\text{End-ADS}}=\int\frac{1}{2}\mathbf{p}^T\mathbf{M}^{-1}\mathbf{p}\rho_{\mathbf{p}}^{\text{End-ADS}}\left(\mathbf{p}\right)d\mathbf{p} \ . \tag{52}$$

Substituting Eq. (49) into the RHS of Eq. (52) and performing the integral, we obtain

$$\frac{1}{2}\left\langle\mathbf{p}^T\mathbf{M}^{-1}\mathbf{p}\right\rangle_{\text{End-ADS}}=e^{-\nu\Delta t}\frac{1}{2}\left\langle\mathbf{p}^T\mathbf{M}^{-1}\mathbf{p}\right\rangle_{\text{Side-ADS}}+\left(1-e^{-\nu\Delta t}\right)\frac{3N}{2\beta} \ , \tag{53}$$

or equivalently

$$\left|\frac{1}{2}\left\langle\mathbf{p}^T\mathbf{M}^{-1}\mathbf{p}\right\rangle_{\text{End-ADS}}-\frac{3N}{2\beta}\right|=e^{-\nu\Delta t}\left|\frac{1}{2}\left\langle\mathbf{p}^T\mathbf{M}^{-1}\mathbf{p}\right\rangle_{\text{Side-ADS}}-\frac{3N}{2\beta}\right| \ . \tag{54}$$

Because the exact value of the averaged kinetic energy is $3N/2\beta$, Eq. (54) suggests that the averaged kinetic energy produced by 'end-Andersen' is more accurate than that produced by 'side-Andersen'.

2) Langevin dynamics

Note that Eq. (6) is an exact solution for the OU process for a finite time interval $\Delta t$. It is trivial to verify that Eq. (41) also holds for Langevin dynamics. This suggests that the stationary state distribution of 'side-Andersen' and that of 'end-Andersen' have the relation

$$\rho^{\text{End-Lang}}\left(\mathbf{x},\mathbf{p}\right)=e^{\mathcal{L}_T\Delta t/2}\rho^{\text{Side-Lang}}\left(\mathbf{x},\mathbf{p}\right) \quad . \tag{55}$$

Since $e^{\mathcal{L}_T\Delta t/2}$ in the Langevin thermostat does *not* change the marginal distribution of $\mathbf{x}$, 'end-Langevin' and 'side-Langevin' share the same stationary state marginal distribution of the coordinate

$$\rho_{\mathbf{x}}^{\text{End-Lang}}\left(\mathbf{x}\right)=\rho_{\mathbf{x}}^{\text{Side-Lang}}\left(\mathbf{x}\right) \quad . \tag{56}$$

Integration over $\mathbf{x}$ in Eq. (55) produces

$$\rho_{\mathbf{p}}^{\text{End-Lang}}\left(\mathbf{p}\right)=e^{\mathcal{L}_T\Delta t/2}\rho_{\mathbf{p}}^{\text{Side-Lang}}\left(\mathbf{p}\right) \tag{57}$$

or equivalently

$$\begin{aligned}\rho_{\mathbf{p}}^{\text{End-Lang}}\left(\mathbf{p}\right)=&\left[\frac{\beta}{2\pi\left(1-e^{-\gamma\Delta t}\right)}\right]^{3N/2}\left|\mathbf{M}\right|^{-1/2}\int d\mathbf{p}_0\rho_{\mathbf{p}}^{\text{Side-Lang}}\left(\mathbf{p}_0\right)\\&\times\exp\left[-\frac{\beta}{2\left(1-e^{-\gamma\Delta t}\right)}\left(\mathbf{p}-e^{-\gamma\Delta t/2}\mathbf{p}_0\right)^T\mathbf{M}^{-1}\left(\mathbf{p}-e^{-\gamma\Delta t/2}\mathbf{p}_0\right)\right]\end{aligned} \quad . \tag{58}$$

It is easy to verify that the difference between the marginal distribution of the momentum in Eq. (58) and the Maxwell momentum distribution $\rho_{\mathrm{MB}}(\mathbf{p})$ [Eq. (14)] is

$$\begin{aligned}\rho_{\mathbf{p}}^{\text{End-Lang}}(\mathbf{p})-\rho_{\mathrm{MB}}(\mathbf{p})=&\left[\frac{\beta}{2\pi\left(1-e^{-\gamma\Delta t}\right)}\right]^{3N/2}|\mathbf{M}|^{-1/2}\int d\mathbf{p}_0\left[\rho_{\mathbf{p}}^{\text{Side-Lang}}(\mathbf{p}_0)\right.\\&\left.-\rho_{\mathrm{MB}}(\mathbf{p}_0)\right]\exp\left[-\frac{\beta}{2\left(1-e^{-\gamma\Delta t}\right)}\left(\mathbf{p}-e^{-\gamma\Delta t/2}\mathbf{p}_0\right)^T\mathbf{M}^{-1}\left(\mathbf{p}-e^{-\gamma\Delta t/2}\mathbf{p}_0\right)\right]\end{aligned}\quad . \tag{59}$$

Consider the absolute value $\left|\rho_{\mathbf{p}}^{\text{End-Lang}}(\mathbf{p})-\rho_{\mathrm{MB}}(\mathbf{p})\right|$. Eq. (59) leads to the inequality

$$\begin{aligned}\left|\rho_{\mathbf{p}}^{\text{End-Lang}}(\mathbf{p})-\rho_{\mathrm{MB}}(\mathbf{p})\right|\le&\left[\frac{\beta}{2\pi\left(1-e^{-\gamma\Delta t}\right)}\right]^{3N/2}|\mathbf{M}|^{-1/2}\int d\mathbf{p}_0\left|\rho_{\mathbf{p}}^{\text{Side-Lang}}(\mathbf{p}_0)\right.\\&\left.-\rho_{\mathrm{MB}}(\mathbf{p}_0)\right|\exp\left[-\frac{\beta}{2\left(1-e^{-\gamma\Delta t}\right)}\left(\mathbf{p}-e^{-\gamma\Delta t/2}\mathbf{p}_0\right)^T\mathbf{M}^{-1}\left(\mathbf{p}-e^{-\gamma\Delta t/2}\mathbf{p}_0\right)\right]\end{aligned}\quad , \tag{60}$$

where the equality holds if and only if $\rho_{\mathbf{p}}^{\text{Side-Lang}}(\mathbf{p})\equiv\rho_{\mathrm{MB}}(\mathbf{p})$. Integration of Eq. (60) over $\mathbf{p}$ produces

$$\begin{aligned}\int\left|\rho_{\mathbf{p}}^{\text{End-Lang}}(\mathbf{p})-\rho_{\mathrm{MB}}(\mathbf{p})\right|d\mathbf{p}\le&\left[\frac{\beta}{2\pi\left(1-e^{-\gamma\Delta t}\right)}\right]^{3N/2}|\mathbf{M}|^{-1/2}\\&\times\int d\mathbf{p}d\mathbf{p}_0\left|\rho_{\mathbf{p}}^{\text{Side-Lang}}(\mathbf{p}_0)-\rho_{\mathrm{MB}}(\mathbf{p}_0)\right|\\&\times\exp\left[-\frac{\beta}{2\left(1-e^{-\gamma\Delta t}\right)}\left(\mathbf{p}-e^{-\gamma\Delta t/2}\mathbf{p}_0\right)^T\mathbf{M}^{-1}\left(\mathbf{p}-e^{-\gamma\Delta t/2}\mathbf{p}_0\right)\right]\end{aligned}\quad . \tag{61}$$

Integration over $\mathbf{p}$ in the RHS of Eq. (61) leads to the following inequality

$$\int\left|\rho_{\mathbf{p}}^{\text{End-Lang}}(\mathbf{p})-\rho_{\mathrm{MB}}(\mathbf{p})\right|d\mathbf{p}\le\int\left|\rho_{\mathbf{p}}^{\text{Side-Lang}}(\mathbf{p})-\rho_{\mathrm{MB}}(\mathbf{p})\right|d\mathbf{p}\ , \tag{62}$$

where the equality holds if and only if $\rho_{\mathbf{p}}^{\text{Side-Lang}}(\mathbf{p}) \equiv \rho_{\text{MB}}(\mathbf{p})$. That is, the stationary state marginal distribution of the momentum produced by 'end-Langevin' is not less accurate than that obtained by 'side-Langevin'. (Here we consider the absolute-value norm or the $L^1$ norm of the difference between the Maxwell momentum distribution and the stationary state marginal distribution of the momentum.)

Consider the averaged kinetic energy produced by 'end-Langevin'

$$\frac{1}{2}\left\langle \mathbf{p}^T \mathbf{M}^{-1} \mathbf{p} \right\rangle_{\text{End-Lang}} = \int \frac{1}{2} \mathbf{p}^T \mathbf{M}^{-1} \mathbf{p} \rho_{\mathbf{p}}^{\text{End-Lang}}(\mathbf{p}) d\mathbf{p} \quad . \tag{63}$$

Substituting Eq. (58) into Eq. (63) and performing the integral, we obtain

$$\frac{1}{2}\left\langle \mathbf{p}^T \mathbf{M}^{-1} \mathbf{p} \right\rangle_{\text{End-Lang}} = e^{-\gamma \Delta t} \frac{1}{2}\left\langle \mathbf{p}^T \mathbf{M}^{-1} \mathbf{p} \right\rangle_{\text{Side-Lang}} + \left(1 - e^{-\gamma \Delta t}\right) \frac{3N}{2\beta} \quad , \tag{64}$$

or equivalently

$$\left| \frac{1}{2}\left\langle \mathbf{p}^T \mathbf{M}^{-1} \mathbf{p} \right\rangle_{\text{End-Lang}} - \frac{3N}{2\beta} \right| = e^{-\gamma \Delta t} \left| \frac{1}{2}\left\langle \mathbf{p}^T \mathbf{M}^{-1} \mathbf{p} \right\rangle_{\text{Side-Lang}} - \frac{3N}{2\beta} \right| \quad . \tag{65}$$

That is, the averaged kinetic energy produced by 'end-Langevin' is in principle more accurate than that produced by 'side-Langevin'.

In summary, when either the 'side' or 'end' scheme is employed, as long as the thermostat process maintains the Maxwell momentum distribution even when $\Delta t$ is finite, the exact momentum distribution is approached in the harmonic limit, regardless of the time interval $\Delta t$ (as long as the matrix $\mathbf{1} - \mathbf{A}\mathbf{M}^{-1} \frac{\Delta t^2}{4}$ is positive-definite). More interestingly, when such a thermostat process is applied to a general system, it is proved that both the 'side' and 'end' schemes lead to the same configurational distribution, while the 'end' scheme in principle produces a more accurate momentum distribution than the 'side' scheme does. As the 'side' scheme is more symmetrized than the 'end' scheme, one would expect that the former should perform better than the latter. Our analysis, however, reveals that the 'end' scheme is superior to the 'side' scheme in sampling the whole phase space.

The same conclusions could be drawn for NHC when the numerical solution for the exact phase space propagator $e^{\mathcal{L}_T \Delta t}$ for the NHC thermostat part [Eq. (138)] is effectively accurate. This is also verified by the numerical examples in next section.

## V. Numerical Examples

### V-1. Classical canonical ensembles *via* MD

#### 1. Simulation detail

We perform numerical tests for several typical systems. The two 1-dimensional models are a harmonic potential $U(x) = m\omega^2 x^2/2$ (with the mass $m=1$ and the frequency $\omega=1$) for the inverse temperature $\beta = 8$ and a quartic potential $U(x) = x^4/4$ (with the mass $m=1$) for $\beta = 8$. Note that the second model contains no harmonic term. So it presents a good example to test numerical behaviors of an algorithm in the anharmonic region.

Three typical realistic systems are also investigated. The first example is the $H_2O$ molecule with the accurate potential energy surface developed by Partridge and Schwenke from extensive *ab initio* calculations and experimental data[48]. As the explicit form of the PES is available, that of the force can be expressed. The MD simulations are performed for $T = 100\,\text{K}$. The time interval ranges from ~0.24 fs to ~2.66 fs (10 ~ 110 au) or to the value that breaks down the propagation of the thermostat. After equilibrating the system, 20 trajectories with each propagated up to ~1.2 ns are used for estimating the energies. The second molecular system is $(Ne)_{13}$, a Lennard-Jones (LJ) cluster. The parameters of the system are described in Ref.[49]. The MD simulations are performed for $T = 14\,\mathrm{K}$. The time interval ranges from 1 fs to 80~82 fs. After the system is equilibrated, 20 trajectories with each propagated up to ~1 ns are used for estimating the energies. The third example is liquid water, a condensed phase system. We employ the POLI2VS–a flexible, polarizable-type force field for liquid water developed by Hasegawa and Tanimura[50]. MD simulations are carried out at $T = 298.15$ K with the liquid density $\rho_l = 0.997\,\mathrm{g\cdot cm^{-3}}$ for a system of 216 water molecules in a box with periodic boundary conditions applied using the minimum image convention. After equilibrating the system, 20 MD trajectories with each propagated up to ~100 ps are used for estimating thermodynamic properties. The time interval is from 0.1 fs to 1.6 fs or to the value that breaks down the propagation of the thermostat.

Both the average potential energy and the average kinetic energy are computed[39]. Each of these thermodynamic properties is plotted as a function of the time interval $\Delta t$. In principle, as $\Delta t$ is small enough, the same converged results should be obtained for all schemes and for all thermostats.

## 2. Results and discussions

### a) Comparison between the 'side' and 'end' schemes

We first compare the performance of the 'side' scheme to that of the 'end' scheme, where the Andersen thermostat, Langevin dynamics, and NHC are employed as the thermostats. We study classical canonical ensembles *via* MD. The first three systems (the harmonic oscillator, the quartic potential, and the $H_2O$ molecule) are employed for demonstration. While the MD results for the average potential energy are shown in Fig. 1, those for the average kinetic energy are depicted in Fig. 2. The MD results in Figs. 1 and 2 are consistent with our analytical analysis for the 'side' and 'end' schemes in Section IV. For the harmonic system both the 'side' and 'end' schemes produce the same results for either the kinetic or potential energy. For general systems the 'end' scheme leads to more accurate results for the kinetic energy than the 'side' scheme does, while both schemes produce the same results for the potential energy, irrespective of which type of thermostat is employed. The numerical results in Figs. 1-2 agree well with our analysis presented in Section IV.

Because the 'side' and 'end' schemes in principle generate the same configurational distribution, below we only compare the 'side' and 'middle' schemes.

**b) Comparison between the 'side' and 'middle' schemes**

We first study the two 1-dimensional models. Fig. 3a compares the algorithms for the 1-dimensional harmonic potential. In agreement with our previous analysis in the harmonic limit, the 'middle' scheme produces accurate average potential energy value that is insensitive to the time interval $\Delta t$, while the 'side' scheme does progressively worse as $\Delta t$ increases. Fig. 3b then depicts the results for the 1-dimensional quartic potential. It also shows that the 'middle' scheme is more accurate and more robust than the 'side' one, regardless of which type of thermostat is employed.

We then investigate the three typical molecular systems. The first system is the $H_2O$ molecule. Fig. 4a shows that all algorithms approach the same results as the time interval is decreased. This agrees with the fact that the algorithms are in principle equivalent as the time interval approaches zero. The fully converged result is obtained at $\Delta t = 0.24\,\text{fs}$. As the time interval increases, the deviation from the converged result for the 'side' scheme is about an order of magnitude (or more) larger than that for the 'middle' scheme, regardless of which type of thermostat is used. The absolute deviation of the average potential energy per atom $\langle U(\mathbf{x})\rangle / (N_{\text{atom}} k_B)$ for the 'middle' scheme is less than $\sim 0.05\,\text{K}$ at $\Delta t = 0.48\,\text{fs}$ and less than $0.27\,\text{K}$ at $\Delta t = 2.18\,\text{fs}$. For comparison, the same property for the 'side' scheme increases from $\sim 0.9\,\text{K}$ at $\Delta t = 0.48\,\text{fs}$ to more than $\sim 63\,\text{K}$ at $\Delta t = 2.18\,\text{fs}$. The three types of thermostats produce similar results in either scheme.

The second molecular system is the cluster $(Ne)_{13}$. Fig. 4b depicts performances of different integrators for simulating $(Ne)_{13}$. All the integrators approach to one another as the time interval decreases. While the absolute deviation of the average potential energy per atom $\langle U(\mathbf{x})\rangle / (N_{\text{atom}} k_B)$ from the converged result for the 'middle' scheme is $\sim 0.04\,\text{K}$ at $\Delta t = 30\,\text{fs}$ and $\sim 0.14\,\text{K}$ at $\Delta t = 70\,\text{fs}$, that for the 'side' scheme is $\sim 0.18\,\text{K}$ at $\Delta t = 30\,\text{fs}$ and $\sim 1\,\text{K}$ at $\Delta t = 70\,\text{fs}$.

The third example is liquid water. As presented in Fig. 4c, all integrators lead to the same converged result (within the statistical error) when the time interval $\Delta t \le 0.2\,\text{fs}$. The 'middle' scheme is more robust than the 'side' one. While the 'side' scheme fails when the time interval $\Delta t$ is greater than ~1.46 fs, the 'middle' scheme still performs well until $\sim 1.6\,\text{fs}$. The absolute deviation of the average potential energy per atom $\langle U(\mathbf{x})\rangle / (N_{\text{atom}} k_B)$ produced by the 'side' scheme is as large as $\sim 34\,\mathrm{K}$ at the time interval $\Delta t = 1.4\ \mathrm{fs}$. For comparison, the same property calculated by the 'middle' scheme is ~2 K at $\Delta t = 1.4\ \mathrm{fs}$ and less than ~2.7 K at $\Delta t \sim 1.6\ \mathrm{fs}$.

While Figs. 3 and 4 of the paper demonstrate the MD results for the average potential energy for the five systems, Figs. 5 and 6 show the MD results for the average kinetic energy for the same systems. While the 'middle' scheme is superior to the 'side' scheme in sampling the coordinate space, the momentum distribution produced by the 'middle' scheme is less accurate than that obtained by the 'side' scheme.

The results in Figs. 1-6 suggest that the 'end' scheme is the best of the three ones for sampling the momentum space while the 'middle' scheme demonstrates the best performance for sampling the coordinate space.

### V-2. Quantum canonical ensembles *via* PIMD

As discussed in Appendix D, all thermodynamic properties depend on the configurational sampling of the path integral beads in the PIMD simulations. Because the 'side' and 'end' schemes in principle generate the same configurational distribution, both schemes in principle produce the same results for any thermodynamic properties for quantum canonical ensembles. So we only compare the 'side' and 'middle' schemes.

We apply the two schemes to PIMD simulations for studying the (quantum) canonical ensemble for liquid water at the state point $T = 298.15$ K and $\rho_l = 0.997\,\mathrm{g \cdot cm^{-3}}$. The same force field (POLI2VS) is used[50]. $P = 48$ path integral beads are employed for simulating 216 water molecules in a box with periodic boundary conditions applied using the minimum image convention. After the system approaches equilibrium, 8 PIMD trajectories with each propagated up to $\sim 50\,\mathrm{ps}$ are used to evaluating thermodynamic properties. The time interval for PIMD ranges from 0.1 fs to 0.75 fs or to the value that breaks down the propagation of the thermostat. The staging transformation[45, 46, 51] of path integral beads is employed.

As presented in Fig. 7a, all algorithms (in the two schemes) lead to nearly the same result for the primitive estimator for the average kinetic energy per atom at the time interval $\Delta t = 0.1\,\mathrm{fs}$. As suggested in Ref. [24], the difference $\Delta E_{\mathrm{kin}}$ between the result of the primitive estimator and that of the virial estimator[52] is a reasonable quantity for measuring the behavior of the PIMD integrator. Fig. 7b shows that the difference $\Delta E_{\mathrm{kin}}$ is close to zero at $\Delta t = 0.1\,\mathrm{fs}$. While the difference $\Delta E_{\mathrm{kin}}$ for the 'middle' scheme is less than 0.8 K at $\Delta t = 0.2\,\mathrm{fs}$ and less than $1.4\,\mathrm{K}$ at $\Delta t = 0.6\,\mathrm{fs}$, that for the 'side' scheme is already larger than $19\,\mathrm{K}$ at $\Delta t = 0.2\,\mathrm{fs}$ and around ~180 K at $\Delta t = 0.6\,\mathrm{fs}$. Fig. 7c demonstrates that the average potential energy per atom obtained by the 'middle' scheme agrees well with that produced by the 'side' scheme within the statistical error bar at the time interval $\Delta t = 0.1\,\mathrm{fs}$, regardless of which thermostat is used. While the absolute deviation (from the converged result at $\Delta t = 0.1\,\mathrm{fs}$) for the 'middle' scheme is less than ~3 K at $\Delta t = 0.6\,\mathrm{fs}$, that for the 'side' scheme is already greater than 37 K at $\Delta t = 0.6\,\mathrm{fs}$. Comparing to the 'side' scheme, the 'middle' scheme reduces the error by about an order of magnitude for the same time interval.

## VI. Conclusion remarks

As demonstrated in Figs. 1-7 for the numerical tests in the two 1-dimensional models and three typical realistic molecular systems that range from the gas phase, clusters, to the condensed phase, different thermostats show similar numerical performance behaviors in evaluating thermodynamic properties when the same scheme is applied. The three typical thermostats (the Andersen thermostat, Langevin dynamics, and NHC) are comparable to one another when the same scheme is employed. The conclusion may be generalized to other types of thermostat for the canonical ensemble.

It is then often a matter of taste or of convenience to choose a type of thermostat in a simulation. While such as the Andersen thermostat and Langevin dynamics are stochastic, such as NHC is deterministic and time-reversible. Although all algorithms in principle lead to the same converged results as the time interval $\Delta t$ approaches zero, the scheme of choice is particularly important in terms of accuracy as a function of the (finite) time interval. The average kinetic energy (per degree of freedom) is often used for estimating how well the temperature is controlled by the thermostat algorithm, i.e., $\langle \mathbf{p}^T \mathbf{M}^{-1} \mathbf{p} \rangle / 3Nk_{\mathrm{B}} = T$. In this regard, the 'side' or 'end' scheme seems to perform well in controlling the temperature in the simulation. This is perhaps why the 'side' or 'end' scheme has earlier been implemented in many different thermostat algorithms. While the 'middle' scheme appears to do worse in controlling the temperature in the simulation, it actually performs better for configurational sampling for the canonical ensemble—it increases the time interval of the propagation by from a factor of 4~5 to an order of magnitude for achieving the same accuracy. Because most thermodynamic properties depend on configurational sampling in MD simulations and all thermodynamic properties do so in PIMD simulations, the 'middle' scheme [Eq. (29)] offers a simple, robust, efficient, and accurate approach for a thermostat, regardless of whether it is stochastic or deterministic. That is, the original work on Langevin dynamics for MD[20] and that for PIMD[24] may be generalized to other types of thermostat.

In summary, we suggest that the ‘middle’ scheme should be considered for use in MD and PIMD simulations for canonical ensembles (and even more generally, isothermal-isobaric ensembles, grand canonical ensembles, *etc*.), regardless of which type of thermostat is preferred to implement. Since it is straightforward to integrate the code for the ‘middle’ scheme for any typical thermostats in simulation packages, we expect that the results that we present in the paper will encourage others to use the ‘middle’ scheme as well to study systems of their interest.

Finally, we note that in the paper we have not used any multiple time-scale techniques for physical degrees of freedom. Multiple time-scale techniques may certainly be employed for the physical degrees of freedom in all the schemes when it improve the efficiency while not losing much accuracy. The ‘middle’ scheme is still expected to perform better than other schemes for configurational sampling. (For instance, it has already been demonstrated when RESPA[13] is used for Eq. (172) for PIMD[24].) We also note that some more sophisticated thermostats with isokinetic constraints [e.g., the isokinetic Nosé-Hoover RESPA (INR) method[53, 54], Nose-Hoover-Langevin (NHL) method[55], stochastic-isokinetic Nosé-Hoover RESPA (SIN(R)) method[56]] have been recently developed, especially with the multiple time-scale technique such as RESPA[13] for physical degrees of freedom of systems that have different time scales. We also note that in the paper we have not used such as the SHAKE[57]/RATTLE[58] algorithms for systems with constraints for bond lengths or angles, for which additional care should be taken care of (e.g., see Ref. [59]). It will certainly be interesting to investigate the ‘middle’ scheme and other ones with such as holonomic and/or isokinetic constraints in future work[60].

## SUPPLEMENTARY MATERIAL

See supplementary material for more discussion on Appendix A and on optimal thermostat parameters.

## ACKNOWLEDGMENT

This work was supported by the Ministry of Science and Technology of China (MOST) Grant No. 2016YFC0202803, by the National Science Foundation of China (NSFC) Grants No. 21373018 and No. 21573007, by the Recruitment Program of Global Experts, by Specialized Research Fund for the Doctoral Program of Higher Education No. 20130001110009, and by Special Program for Applied Research on Super Computation of the NSFC-Guangdong Joint Fund (the second phase). We acknowledge the Beijing and Tianjin supercomputer centers for providing computational resources.

## Appendix A. Stationary state distribution of the Andersen thermostat for a finite time interval

Consider a 1-dimensional harmonic system where Eq. (30) becomes

$$U(x) = A\left(x - x_{eq}\right)^2 / 2 \ . \tag{66}$$

When the time interval $\Delta t$ is finite, the full Kolmogorov operator for the Andersen thermostat is broken down into three parts $\mathcal{L} = \mathcal{L}_p + \mathcal{L}_x + \mathcal{L}_T$ with

$$\mathcal{L}_p \rho = A\left(x - x_{eq}\right)\frac{\partial \rho}{\partial p} \ , \tag{67}$$

$$\mathcal{L}_x \rho = -\frac{p}{m}\frac{\partial \rho}{\partial x} \ , \tag{68}$$

and

$$\mathcal{L}_T \rho = \nu\left[\rho_{\mathrm{MB}}(p)\int \rho dp - \rho\right] \ . \tag{69}$$

Use the 'middle' scheme [Eq. (29)] as the example. Define the following densities

$$\begin{aligned}
\rho_{n,0}(x,p) &\equiv \left(e^{\mathcal{L}^{\mathrm{Middle}}\Delta t}\right)^n \rho_0(x,p) \\
\rho_{n,1}(x,p) &\equiv e^{\mathcal{L}_p \Delta t/2}\rho_{n,0}(x,p) \\
\rho_{n,2}(x,p) &\equiv e^{\mathcal{L}_x \Delta t/2}\rho_{n,1}(x,p) \\
\rho_{n,3}(x,p) &\equiv e^{\mathcal{L}_T \Delta t}\rho_{n,2}(x,p) \\
\rho_{n,4}(x,p) &\equiv e^{\mathcal{L}_x \Delta t/2}\rho_{n,3}(x,p)
\end{aligned} \quad , \tag{70}$$

which leads to

$$\rho_{n+1,0}(x,p) = e^{\mathcal{L}_p \Delta t/2}\rho_{n,4}(x,p) \ . \tag{71}$$

We introduce the notation

$$\langle O \rangle_{n,i} \equiv \int \rho_{n,i}(x,p) O(x,p)\, dxdp \quad i = 0,\cdots,4 \ , \tag{72}$$

where $O(x,p)$ is a physical property of interest. For example, the mean coordinate displacement and the mean momentum can be expressed as

$$\boldsymbol{\zeta}_{n+1,0}^{(1)} \equiv \begin{pmatrix} \langle x - x_{eq} \rangle_{n+1,0} \\ \langle p \rangle_{n+1,0} \end{pmatrix} = \begin{pmatrix} \int \rho_{n+1,0}(x,p)(x - x_{eq})\, dxdp \\ \int \rho_{n+1,0}(x,p)\, p\, dxdp \end{pmatrix} . \tag{73}$$

Substituting Eq. (71) into Eq. (73) and then performing the integral leads to

$$\boldsymbol{\zeta}_{n+1,0}^{(1)} = \begin{pmatrix} \langle x - x_{eq} \rangle_{n,4} \\ \langle p \rangle_{n,4} - A\dfrac{\Delta t}{2} \langle x - x_{eq} \rangle_{n,4} \end{pmatrix} = \mathbf{A}_1^{(1)} \boldsymbol{\zeta}_{n,4}^{(1)} \ , \tag{74}$$

with

$$\mathbf{A}_1^{(1)} = \begin{pmatrix} 1 & 0 \\ -A\dfrac{\Delta t}{2} & 1 \end{pmatrix} . \tag{75}$$

Similarly, one could obtain

$$\begin{aligned} \boldsymbol{\zeta}_{n,4}^{(1)} &= \mathbf{A}_2^{(1)} \boldsymbol{\zeta}_{n,3}^{(1)} \\ \boldsymbol{\zeta}_{n,3}^{(1)} &= \mathbf{A}_3^{(1)} \boldsymbol{\zeta}_{n,2}^{(1)} \\ \boldsymbol{\zeta}_{n,2}^{(1)} &= \mathbf{A}_2^{(1)} \boldsymbol{\zeta}_{n,1}^{(1)} \\ \boldsymbol{\zeta}_{n,1}^{(1)} &= \mathbf{A}_1^{(1)} \boldsymbol{\zeta}_{n,0}^{(1)} \end{aligned} \tag{76}$$

with

$$\mathbf{A}_2^{(1)} = \begin{pmatrix} 1 & \dfrac{\Delta t}{2m} \\ 0 & 1 \end{pmatrix} \tag{77}$$

$$\mathbf{A}_3^{(1)} = \begin{pmatrix} 1 & 0 \\ 0 & e^{-\nu\Delta t} \end{pmatrix} \quad . \tag{78}$$

Define $\tilde{\mathbf{A}}_1 = \mathbf{A}_1^{(1)}\mathbf{A}_2^{(1)}\mathbf{A}_3^{(1)}\mathbf{A}_2^{(1)}\mathbf{A}_1^{(1)}$. Eq. (74) and Eq. (76) then leads to

$$\boldsymbol{\zeta}_{n+1,0}^{(1)} = \tilde{\mathbf{A}}_1 \boldsymbol{\zeta}_{n,0}^{(1)} \ , \tag{79}$$

or equivalently

$$\boldsymbol{\zeta}_{n,0}^{(1)} = \tilde{\mathbf{A}}_1^n \boldsymbol{\zeta}_{0,0}^{(1)} \ . \tag{80}$$

Because the spectral radius of matrix $\tilde{\mathbf{A}}_1$ is less than 1[61], we have

$$\boldsymbol{\zeta}_{n,0}^{(1)} \rightarrow \mathbf{0} \ \text{ as } \ n \rightarrow \infty. \tag{81}$$

Analogously, the evolution of the second-order moment vector

$$\boldsymbol{\zeta}_{n,0}^{(2)} \equiv \left( \left\langle \left(x - x_{eq}\right)^2 \right\rangle_{n,0}, \left\langle \left(x - x_{eq}\right) p \right\rangle_{n,0}, \left\langle p^2 \right\rangle_{n,0} \right)^T \tag{82}$$

satisfies

$$\begin{aligned} \boldsymbol{\zeta}_{n+1,0}^{(2)} &= \mathbf{A}_1^{(2)} \boldsymbol{\zeta}_{n,4}^{(2)} \\ \boldsymbol{\zeta}_{n,4}^{(2)} &= \mathbf{A}_2^{(2)} \boldsymbol{\zeta}_{n,3}^{(2)} \\ \boldsymbol{\zeta}_{n,3}^{(2)} &= \mathbf{A}_3^{(2)} \boldsymbol{\zeta}_{n,2}^{(2)} + \mathbf{b}_2 \\ \boldsymbol{\zeta}_{n,2}^{(2)} &= \mathbf{A}_2^{(2)} \boldsymbol{\zeta}_{n,1}^{(2)} \\ \boldsymbol{\zeta}_{n,1}^{(2)} &= \mathbf{A}_1^{(2)} \boldsymbol{\zeta}_{n,0}^{(2)} \end{aligned} \tag{83}$$

where

$$\mathbf{A}_1^{(2)} = \begin{pmatrix} 1 & 0 & 0 \\ -A\frac{\Delta t}{2} & 1 & 0 \\ A^2\frac{\Delta t^2}{4} & -A\Delta t & 1 \end{pmatrix}, \quad \mathbf{A}_2^{(2)} = \begin{pmatrix} 1 & \frac{\Delta t}{m} & \frac{\Delta t^2}{4m^2} \\ 0 & 1 & \frac{\Delta t}{2m} \\ 0 & 0 & 1 \end{pmatrix}, \quad \mathbf{A}_3^{(2)} = \begin{pmatrix} 1 & 0 & 0 \\ 0 & e^{-\nu\Delta t} & 0 \\ 0 & 0 & e^{-\nu\Delta t} \end{pmatrix}. \tag{84}$$

$$\mathbf{b}_2 = \left(0, 0, \frac{m}{\beta}\left(1 - e^{-\nu\Delta t}\right)\right)^T$$

Define $\tilde{\mathbf{A}}_2 = \mathbf{A}_1^{(2)}\mathbf{A}_2^{(2)}\mathbf{A}_3^{(2)}\mathbf{A}_2^{(2)}\mathbf{A}_1^{(2)}$ and $\tilde{\mathbf{b}}_2 = \mathbf{A}_1^{(2)}\mathbf{A}_2^{(2)}\mathbf{b}_2$ so that we obtain

$$\zeta_{n+1,0}^{(2)} = \tilde{\mathbf{A}}_2\zeta_{n,0}^{(2)} + \tilde{\mathbf{b}}_2 \ . \tag{85}$$

Rearranging Eq. (85) leads to

$$\zeta_{n+1,0}^{(2)} - \left(\mathbf{I} - \tilde{\mathbf{A}}_2\right)^{-1}\tilde{\mathbf{b}}_2 = \tilde{\mathbf{A}}_2\left[\zeta_{n,0}^{(2)} - \left(\mathbf{I} - \tilde{\mathbf{A}}_2\right)^{-1}\tilde{\mathbf{b}}_2\right], \tag{86}$$

or equivalently,

$$\zeta_{n,0}^{(2)} - \left(\mathbf{I} - \tilde{\mathbf{A}}_2\right)^{-1}\tilde{\mathbf{b}}_2 = \tilde{\mathbf{A}}_2^n\left[\zeta_{0,0}^{(2)} - \left(\mathbf{I} - \tilde{\mathbf{A}}_2\right)^{-1}\tilde{\mathbf{b}}_2\right]. \tag{87}$$

It is easy to verify that the spectral radius of matrix $\tilde{\mathbf{A}}_2$ is less than 1[61], so we may show $\zeta_{n,0}^{(2)} \to \left(\mathbf{I} - \tilde{\mathbf{A}}_2\right)^{-1}\tilde{\mathbf{b}}_2$ as $n \to \infty$. Using Eq. (84) we obtain the explicit expression

$$\zeta_{n,0}^{(2)} \to \left(\frac{1}{\beta A}, 0, \frac{m}{\beta} - \frac{A\Delta t^2}{4\beta}\right)^T \quad \text{as } n \to \infty. \ . \tag{88}$$

With the mean value [Eq. (81)] and the second-order moments [Eq. (88)], it is not sufficient to obtain the stationary state distribution. Higher-order moments are necessary. We denote the *k*-th order moment vector as $\boldsymbol{\zeta}_{n,0}^{(k)} \equiv \left( \left\langle \left( x - x_{eq} \right)^{k-j} p^j \right\rangle_{n,0}, j = 0, \cdots, k \right)^T$, the evolution of which satisfies

$$\boldsymbol{\zeta}_{n+1,0}^{(k)} = \mathbf{A}_1^{(k)} \mathbf{A}_2^{(k)} \mathbf{A}_3^{(k)} \mathbf{A}_2^{(k)} \mathbf{A}_1^{(k)} \boldsymbol{\zeta}_{n,0}^{(k)} + \mathbf{A}_1^{(k)} \mathbf{A}_2^{(k)} \mathbf{b}_{k,n} \tag{89}$$

with

$$\mathbf{A}_1^{(k)} = \begin{pmatrix} 1 & & \\ -A\frac{\Delta t}{2} & 1 & \\ \vdots & \ddots & \ddots \\ \left(-A\frac{\Delta t}{2}\right)^k & \cdots & -A\frac{k\Delta t}{2} & 1 \end{pmatrix}, \quad \mathbf{A}_2^{(k)} = \begin{pmatrix} 1 & \frac{k\Delta t}{2m} & \cdots & \left(\frac{\Delta t}{2m}\right)^k \\ & 1 & \ddots & \vdots \\ & & \ddots & \frac{\Delta t}{2m} \\ & & & 1 \end{pmatrix},$$
$$\mathbf{A}_3^{(k)} = \begin{pmatrix} 1 & & & \\ & e^{-\nu\Delta t} & & \\ & & \ddots & \\ & & & e^{-\nu\Delta t} \end{pmatrix}, \quad \mathbf{b}_{k,n} = \left( 0, 0, \left\langle \left( x - x_{eq} \right)^{k-2} \right\rangle_{n,2} \frac{m}{\beta} \left( 1 - e^{-\nu\Delta t} \right), \cdots \right)^T \tag{90}$$

Here $\mathbf{b}_{k,n}$ is related to the lower-order moments $\boldsymbol{\zeta}_{n,2}^{(k-2)}$, $\boldsymbol{\zeta}_{n,2}^{(k-4)}$, and so on. Define $\tilde{\mathbf{A}}_k = \mathbf{A}_1^{(k)} \mathbf{A}_2^{(k)} \mathbf{A}_3^{(k)} \mathbf{A}_2^{(k)} \mathbf{A}_1^{(k)}$ and $\tilde{\mathbf{b}}_{k,n} = \mathbf{A}_1^{(k)} \mathbf{A}_2^{(k)} \mathbf{b}_{k,n}$. The general formula for Eq. (89) is then

$$\boldsymbol{\zeta}_{n,0}^{(k)} = \tilde{\mathbf{A}}_k^n \boldsymbol{\zeta}_{0,0}^{(k)} + \sum_{m=0}^{n-1} \tilde{\mathbf{A}}_k^{n-1-m} \tilde{\mathbf{b}}_{k,m} \quad . \tag{91}$$

Use mathematical induction. Assume that the limits of the lower-order moments $\boldsymbol{\zeta}_{n,2}^{(k-2)}$, $\boldsymbol{\zeta}_{n,2}^{(k-4)}, \cdots$ exist when $n \to \infty$, and

$$\lim_{n\to\infty} \boldsymbol{\zeta}_{n,0}^{(j)} = \begin{cases} \mathbf{0}, \text{ for odd } j \\ \left( \left( \frac{1}{\beta A} \right)^{j/2} (j-1)!!, 0, \cdots, 0, \left( \frac{m}{\beta} - \frac{A\Delta t^2}{4\beta} \right)^{j/2} (j-1)!! \right)^T, \text{ for even } j \end{cases} \tag{92}$$

holds for all $j<k$. So $\bar{\mathbf{b}}_k \equiv \lim_{n\to\infty}\tilde{\mathbf{b}}_{k,n}$ exists and

$$\bar{\mathbf{b}}_k = \begin{cases} \mathbf{0}, \text{ for odd } k \\ \mathbf{A}_1^{(k)}\mathbf{A}_2^{(k)}\left(0,0,(k-3)!!\left(\frac{1}{\beta A}-\frac{\Delta t^2}{4\beta m}\right)^{\frac{k-2}{2}}\frac{m}{\beta}\left(1-e^{-\nu\Delta t}\right),\cdots\right)^T, \text{ for even } k \end{cases} \tag{93}$$

It is straightforward to verify that the spectral radius of matrix $\tilde{\mathbf{A}}_k$ is less than 1. We may then prove that the limit of Eq. (91) exists as $n\to\infty$ and is

$$\lim_{n\to\infty}\zeta_{n,0}^{(k)} = \left(\mathbf{I}-\tilde{\mathbf{A}}_k\right)^{-1}\bar{\mathbf{b}}_k \quad . \tag{94}$$

It is easy to show that

$$\mathbf{y} = \begin{cases} \mathbf{0}, \text{ for odd } k \\ \left(\left(\frac{1}{\beta A}\right)^{k/2}(k-1)!!,0,\cdots,0,\left(\frac{m}{\beta}-\frac{A\Delta t^2}{4\beta}\right)^{k/2}(k-1)!!\right)^T, \text{ for even } k \end{cases} \tag{95}$$

is a solution to linear equations $\left(\mathbf{I}-\tilde{\mathbf{A}}_k\right)\mathbf{y}=\bar{\mathbf{b}}_k$. Because of the non-singularity of matrix $\mathbf{I}-\tilde{\mathbf{A}}_k$, Eq. (95) is its unique solution that is equivalent to $\left(\mathbf{I}-\tilde{\mathbf{A}}_k\right)^{-1}\bar{\mathbf{b}}_k$, which gives the value of the RHS of Eq. (94). In accordance to the principle of induction, Eq. (92) holds for all $j\geq 1$. With all *k*-th moments, it is straightforward to construct the moment generating function

$$\begin{aligned} g(z_1,z_2) &= \lim_{n\to\infty}\left\langle e^{z_1\left(x-x_{eq}\right)+z_2 p}\right\rangle_{n,0} \\ &= \exp\left[\frac{z_1^2}{2\beta A}+\left(\frac{m}{\beta}-\frac{A\Delta t^2}{4\beta}\right)\frac{z_2^2}{2}\right] \end{aligned} \tag{96}$$

which does exist and is equal to that of the Gaussian distribution

$$\rho^{\text{Middle}} = \frac{1}{\bar{Z}_N} \exp\left[ -\frac{\beta}{2} A\left(x - x_{eq}\right)^2 - \frac{\beta p^2}{2m\left(1 - A\Delta t^2 / 4m\right)} \right] \tag{97}$$

where $\bar{Z}_N$ is the normalization constant. That is, Eq. (97) is the stationary state distribution of 'middle-Andersen'.

It is straightforward to follow the same procedure to show that either 'side-Andersen' or 'end-Andersen' produces the stationary state distribution

$$\rho^{\text{Side}} = \rho^{\text{End}} = \frac{1}{Z_N} \exp\left[ -\frac{\beta}{2} A\left(1 - \frac{A}{m}\frac{\Delta t^2}{4}\right)\left(x - x_{eq}\right)^2 - \frac{\beta p^2}{2m} \right] . \tag{98}$$

where $Z_N$ is the normalization constant.

The above procedure may also be used to obtain the stationary state distribution for Langevin dynamics for the one-dimensional harmonic potential [Eq. (66)]. Leimkuhler and Matthews employed a different approach to get the mean and the second-order moments of the stationary state distribution for Langevin dynamics for a one-dimensional harmonic potential when the time interval is finite[22], but they did not compute higher-order moments, neither did they show the form of the stationary state distribution.

## Appendix B. Optimal collision frequency for the Andersen thermostat

Consider $A = m\omega^2$, that is, Eq. (66) becomes

$$U(x) = m\omega^2 \left(x - x_{eq}\right)^2 / 2 \quad . \tag{99}$$

The propagation of the density distribution in the phase space can be expressed by

$$\frac{\partial \rho}{\partial t} = \mathcal{L}\rho = -\frac{p}{m}\frac{\partial \rho}{\partial x} + m\omega^2(x - x_{eq})\frac{\partial \rho}{\partial p} + \nu\left[\rho_{\mathrm{MB}}\left(p\right)\int \rho dp - \rho\right] . \tag{100}$$

Assume that the conditional density distribution function $\rho \equiv \rho\left(x, p; t \middle| x_0, p_0; 0\right)$ is a solution to Eq. (100). Although the explicit expression of $\rho\left(x, p; t \middle| x_0, p_0; 0\right)$ is difficult to obtain, we directly analyze the coordinate displacement square autocorrelation function, which can be expressed by $\rho$ as

$$\left\langle \left(x(0) - x_{eq}\right)^2 \left(x(t) - x_{eq}\right)^2 \right\rangle = \int \rho_0\left(x_0, p_0\right)\rho\left(x_0 - x_{eq}\right)^2\left(x - x_{eq}\right)^2 dx_0 dp_0 dx dp , \tag{101}$$

where the initial condition satisfies the Boltzmann distribution that is a stationary state distribution for Eq. (100), i.e.,

$$\rho_0\left(x_0, p_0\right) = \frac{\beta\omega}{2\pi} e^{-\beta\left[\frac{1}{2}m\omega^2(x_0 - x_{eq})^2 + \frac{p_0^2}{2m}\right]} . \tag{102}$$

Consider the time derivative of Eq. (101), i.e.

$$\frac{\partial}{\partial t}\left\langle \left(x(0) - x_{eq}\right)^2 \left(x(t) - x_{eq}\right)^2 \right\rangle = \int \rho_0\left(x_0, p_0\right)\frac{\partial \rho}{\partial t}\left(x_0 - x_{eq}\right)^2\left(x - x_{eq}\right)^2 dx_0 dp_0 dx dp . \tag{103}$$

Substituting Eq. (100) into Eq. (103) and using integration by parts, we obtain

$$\frac{\partial}{\partial t}\left\langle \left(x(0) - x_{eq}\right)^2 \left(x(t) - x_{eq}\right)^2 \right\rangle = \frac{2}{m}\left\langle \left(x(0) - x_{eq}\right)^2 \left(x(t) - x_{eq}\right) p(t) \right\rangle . \tag{104}$$

Similarly, it is straightforward to verify

$$\begin{aligned}&\frac{\partial}{\partial t}\left\langle \left(x(0)-x_{eq}\right)^2 \left(x(t)-x_{eq}\right) p(t)\right\rangle \\ &=\frac{1}{m}\left\langle \left(x(0)-x_{eq}\right)^2 p^2(t)\right\rangle - m\omega^2 \left\langle \left(x(0)-x_{eq}\right)^2 \left(x(t)-x_{eq}\right)^2\right\rangle \\ &\quad -\nu \left\langle \left(x(0)-x_{eq}\right)^2 x\left((t)-x_{eq}\right) p(t)\right\rangle\end{aligned} \tag{105}$$

and

$$\begin{aligned}&\frac{\partial}{\partial t}\left\langle \left(x(0)-x_{eq}\right)^2 p^2(t)\right\rangle \\ &\quad = -2m\omega^2 \left\langle \left(x(0)-x_{eq}\right)^2 \left(x(t)-x_{eq}\right) p(t)\right\rangle + \frac{\nu}{\beta^2\omega^2} - \nu \left\langle \left(x(0)-x_{eq}\right)^2 p^2(t)\right\rangle\end{aligned} . \tag{106}$$

Eqs. (104)-(106) then form a closed set of first-order linear ODE, expressed in a compact form as

$$\dot{\boldsymbol{\chi}} = \mathbf{A}\boldsymbol{\chi} + \mathbf{b} \ , \tag{107}$$

Where

$$\begin{aligned}\boldsymbol{\chi}(t) &= \left(\chi_1(t), \chi_2(t), \chi_3(t)\right)^T \\ &= \left(\left\langle \left(x(0)-x_{eq}\right)^2 \left(x(t)-x_{eq}\right)^2\right\rangle, \left\langle \left(x(0)-x_{eq}\right)^2 \left(x(t)-x_{eq}\right) p(t)\right\rangle, \left\langle \left(x(0)-x_{eq}\right)^2 p^2(t)\right\rangle\right)^T\end{aligned} , \tag{108}$$

$$\mathbf{b} = \left(0, 0, \frac{\nu}{\beta^2\omega^2}\right)^T \quad , \tag{109}$$

and the linear coefficient matrix

$$\mathbf{A} = \begin{pmatrix} 0 & 2/m & 0 \\ -m\omega^2 & -\nu & 1/m \\ 0 & -2m\omega^2 & -\nu \end{pmatrix} . \tag{110}$$

Solving the ODE [Eq. (107)] with the initial value given by

$$\boldsymbol{\chi}(0) = \left( \frac{3}{\beta^2 m^2 \omega^4}, 0, \frac{1}{\beta^2 \omega^2} \right)^T \ , \tag{111}$$

one obtains

$$\boldsymbol{\chi}(t) = e^{\mathbf{A}t} \left[ \mathbf{A}^{-1}\mathbf{b} + \boldsymbol{\chi}(0) \right] - \mathbf{A}^{-1}\mathbf{b} \ . \tag{112}$$

The characteristic time of the potential energy autocorrelation function

$$\tau_{UU} = \int_0^\infty \frac{\langle U(0)U(t)\rangle - \langle U\rangle^2}{\langle U^2\rangle - \langle U\rangle^2} dt \tag{113}$$

can be shown as

$$\tau_{UU} = \frac{1}{2}\int_0^\infty \left[ \left(\beta m \omega^2\right)^2 \chi_1(t) - 1 \right] dt \ . \tag{114}$$

Substituting Eq. (112) into Eq. (114), we obtain the explicit expression for the characteristic time of the potential autocorrelation function for the one-dimensional harmonic potential [Eq. (99)]

$$\tau_{UU} = \frac{1}{2}\left( \frac{\nu}{\omega^2} + \frac{2}{\nu} \right) \ . \tag{115}$$

The smaller the $\tau_{UU}$ is, the more efficiently the Andersen thermostat explores the potential energy surface and samples the configurational space. When

$$\nu = \nu_{UU}^{(opt)} = \sqrt{2}\omega \ , \tag{116}$$

the characteristic correlation time $\tau_{UU}$ reaches its minimum value

$$\tau_{UU}^{\text{min, ADS}} = \sqrt{2} / \omega \quad . \tag{117}$$

Similarly, the characteristic time of the Hamiltonian autocorrelation function for the one-dimensional harmonic potential [Eq. (99)] may be shown as

$$\tau_{HH} = \frac{\nu}{4\omega^2} + \frac{2}{\nu} \quad . \tag{118}$$

When

$$\nu = \nu_{HH}^{(opt)} = 2\sqrt{2}\omega \quad , \tag{119}$$

$\tau_{HH}$ reaches its minimum value

$$\tau_{HH}^{\text{min, ADS}} = \sqrt{2} / \omega \quad . \tag{120}$$

The procedure above also offers a useful approach to derive the characteristic correlation time for other stochastic thermostats. For instance, in addition to the Andersen thermostat, the approach may be applied to such as Langevin dynamics.

It is interesting to compare the minimum value of the characteristic time of the potential or Hamiltonian autocorrelation function for the Andersen thermostat to that for Langevin dynamics for the harmonic potential [Eq. (99)]. The latter may also be derived from a different approach presented in Appendix A of Ref. [24] or from other different approaches[62, 63]. The minimum characteristic time of the potential or Hamiltonian autocorrelation function for Langevin dynamics[24, 26, 62, 63] is

$$\tau_{UU}^{\text{min, Lang}} = \tau_{HH}^{\text{min, Lang}} = 1 / \omega \quad . \tag{121}$$

The minimum value in Eq. (117) or Eq. (120) for the Andersen thermostat is only $\sqrt{2}$ times of that in Eq. (121) for Langevin dynamics. That is, in terms of sampling efficiency the Andersen thermostat is comparable to Langevin dynamics.

The analysis for the harmonic system may apply to general systems. As demonstrated in numerical examples in Section S2 of the supplementary material[64], for general systems the optimal value of the collision frequency of the Andersen thermostat is about $\sqrt{2}$ times of that of the friction coefficient of Langevin dynamics.

**Appendix C. Numerical algorithms for the thermostats**

**1. Andersen thermostat**

In the conventional algorithm for the Andersen thermostat[5], the collision process is applied after a whole step of the velocity Verlet algorithm is implemented.[3] I.e., the phase space propagator $e^{\mathcal{L}\Delta t}$ employs the splitting in the 'end' scheme [Eq. (28)]. The 'end-Andersen' algorithm for propagating the MD trajectory through a time interval $\Delta t$ is

$$\begin{aligned} &\mathbf{p} \leftarrow \mathbf{p} - U'(\mathbf{x})\frac{\Delta t}{2} \\ &\mathbf{x} \leftarrow \mathbf{x} + \mathbf{M}^{-1}\mathbf{p}\Delta t \quad , \\ &\mathbf{p} \leftarrow \mathbf{p} - U'(\mathbf{x})\frac{\Delta t}{2} \end{aligned} \tag{122}$$

$$\mathbf{p}^{(j)} \leftarrow \sqrt{\frac{1}{\beta}}\mathbf{M}_j^{1/2}\boldsymbol{\theta}_j, \quad \text{if } \mu_j < 1 - e^{-\nu\Delta t} \qquad \left(j = \overline{1, N}\right) . \tag{123}$$

Here $\mathbf{p}^{(j)}$, $\mathbf{M}_j$, $\mu_j$ and $\boldsymbol{\theta}_j$ are the same as discussed for Eq. (12). Note that both $\mu_j$ and $\boldsymbol{\theta}_j$ are different for each invocation of Eq. (123).

When the 'side' scheme [Eq. (27)] is used, the 'side-Andersen' algorithm for propagating the MD trajectory through a time interval $\Delta t$ reads

$$\mathbf{p}^{(j)} \leftarrow \sqrt{\frac{1}{\beta}} \mathbf{M}_j^{1/2} \boldsymbol{\theta}_j, \quad \text{if } \mu_j < 1 - e^{-\nu \Delta t/2} \qquad \left( j = \overline{1, N} \right), \tag{124}$$

$$\begin{aligned} \mathbf{p} &\leftarrow \mathbf{p} - U'(\mathbf{x}) \frac{\Delta t}{2} \\ \mathbf{x} &\leftarrow \mathbf{x} + \mathbf{M}^{-1} \mathbf{p} \Delta t \\ \mathbf{p} &\leftarrow \mathbf{p} - U'(\mathbf{x}) \frac{\Delta t}{2} \end{aligned}, \tag{125}$$

$$\mathbf{p}^{(j)} \leftarrow \sqrt{\frac{1}{\beta}} \mathbf{M}_j^{1/2} \boldsymbol{\theta}_j, \quad \text{if } \mu_j < 1 - e^{-\nu \Delta t/2} \qquad \left( j = \overline{1, N} \right), \tag{126}$$

where $\mu_j$ is a uniformly distributed random number between 0 and 1 and $\boldsymbol{\theta}_j$ is a 3-dimensional Gaussian-distributed random number vector as discussed for Eq. (12). Note that both $\mu_j$ and $\boldsymbol{\theta}_j$ are different for each invocation of Eq. (124) or Eq. (126).

Similarly, when the 'middle' scheme [Eq. (29)] is implemented, the 'middle-Andersen' algorithm for propagating the MD trajectory through a time interval $\Delta t$ is then

$$\begin{aligned} \mathbf{p} &\leftarrow \mathbf{p} - U'(\mathbf{x}) \frac{\Delta t}{2} \\ \mathbf{x} &\leftarrow \mathbf{x} + \mathbf{M}^{-1} \mathbf{p} \frac{\Delta t}{2} \end{aligned}, \tag{127}$$

$$\mathbf{p}^{(j)} \leftarrow \sqrt{\frac{1}{\beta}} \mathbf{M}_j^{1/2} \boldsymbol{\theta}_j, \quad \text{if } \mu_j < 1 - e^{-\nu \Delta t} \qquad \left( j = \overline{1, N} \right), \tag{128}$$

$$\begin{aligned} \mathbf{x} &\leftarrow \mathbf{x} + \mathbf{M}^{-1}\mathbf{p}\frac{\Delta t}{2} \\ \mathbf{p} &\leftarrow \mathbf{p} - U'(\mathbf{x})\frac{\Delta t}{2} \end{aligned} \quad , \tag{129}$$

where both $\mu_j$ and $\boldsymbol{\theta}_j$ are different for each invocation of Eq. (128). Here $\mathbf{p}^{(j)}$, $\mathbf{M}_j$, $\mu_j$ and $\boldsymbol{\theta}_j$ are the same as discussed for Eq. (12).

### 2. Langevin dynamics

The Langevin thermostat algorithm proposed by Bussi *et al.*[25] in 2007 employs the splitting in the 'side' scheme [Eq. (27)]. The 'side-Langevin' algorithm for propagating the MD trajectory through a time interval $\Delta t$ for Eq. (27) becomes

$$\mathbf{p} \leftarrow \tilde{c}_1\mathbf{p} + \tilde{c}_2\sqrt{\frac{1}{\beta}}\mathbf{M}^{1/2}\tilde{\boldsymbol{\eta}} \ , \tag{130}$$

$$\begin{aligned} \mathbf{p} &\leftarrow \mathbf{p} - U'(\mathbf{x})\frac{\Delta t}{2} \\ \mathbf{x} &\leftarrow \mathbf{x} + \mathbf{M}^{-1}\mathbf{p}\Delta t \\ \mathbf{p} &\leftarrow \mathbf{p} - U'(\mathbf{x})\frac{\Delta t}{2} \end{aligned} \quad , \tag{131}$$

$$\mathbf{p} \leftarrow \tilde{c}_1\mathbf{p} + \tilde{c}_2\sqrt{\frac{1}{\beta}}\mathbf{M}^{1/2}\tilde{\boldsymbol{\eta}} \ , \tag{132}$$

where the coefficients $\tilde{c}_1 = e^{-\gamma\Delta t/2}$ and $\tilde{c}_2 = \sqrt{1-\tilde{c}_1^2}$ . $\tilde{\boldsymbol{\eta}}$ is the independent Gaussian-distributed random number vector as discussed for Eq. (2). Note that $\tilde{\boldsymbol{\eta}}$ is different for each invocation of Eq. (130) or Eq. (132).

Similarly, the phase space propagator $e^{\mathcal{L}\Delta t}$ may also use the splitting in the 'end' scheme [Eq. (28)]. The 'end-Langevin' algorithm for propagating the MD trajectory through a time interval $\Delta t$ for Eq. (28) is

$$\begin{aligned}&\mathbf{p}\leftarrow\mathbf{p}-U'(\mathbf{x})\frac{\Delta t}{2}\\&\mathbf{x}\leftarrow\mathbf{x}+\mathbf{M}^{-1}\mathbf{p}\Delta t\\&\mathbf{p}\leftarrow\mathbf{p}-U'(\mathbf{x})\frac{\Delta t}{2}\end{aligned}\qquad(133)$$

$$\mathbf{p}\leftarrow c_1\mathbf{p}+c_2\sqrt{\frac{1}{\beta}}\mathbf{M}^{1/2}\tilde{\boldsymbol{\eta}}\ ,\qquad(134)$$

where the coefficients $c_1=e^{-\gamma\Delta t}$ and $c_2=\sqrt{1-c_1^2}$ . $\tilde{\boldsymbol{\eta}}$ is the independent Gaussian-distributed random number vector as discussed for Eq. (2), which is different for each invocation of Eq. (134).

When the 'middle' scheme [Eq. (29)] is implemented for the phase space propagator, the 'middle-Langevin' algorithm for propagating the MD trajectory through a time interval $\Delta t$ for Eq. (29) reads

$$\begin{aligned}&\mathbf{p}\leftarrow\mathbf{p}-U'(\mathbf{x})\frac{\Delta t}{2}\\&\mathbf{x}\leftarrow\mathbf{x}+\mathbf{M}^{-1}\mathbf{p}\frac{\Delta t}{2}\end{aligned}\ ,\qquad(135)$$

$$\mathbf{p}\leftarrow c_1\mathbf{p}+c_2\sqrt{\frac{1}{\beta}}\mathbf{M}^{1/2}\tilde{\boldsymbol{\eta}}\ ,\qquad(136)$$

$$\begin{aligned} \mathbf{x} &\leftarrow \mathbf{x} + \mathbf{M}^{-1}\mathbf{p}\frac{\Delta t}{2} \\ \mathbf{p} &\leftarrow \mathbf{p} - U'(\mathbf{x})\frac{\Delta t}{2} \end{aligned} \quad , \tag{137}$$

where the coefficients $c_1$ and $c_2$ are the same as those defined in Eq. (134). As used in Eq. (134), the independent Gaussian-distributed random number vector $\tilde{\boldsymbol{\eta}}$ is different for each invocation of Eq. (136). The 'middle-Langevin' algorithm was proposed earlier by Leimkuhler and Matthews[20] and also by Gronbech-Jensen and Farago[21]. Leimkuhler and Matthews have recently suggested that the 'middle-Langevin' is the most efficient Langevin dynamics algorithm for configurational sampling[20, 22] of the canonical ensemble.

### 3. Nosé-Hoover chain

For the equations of motion [Eq. (18)] of NHC , the three relevant Kolmogorov operators are $\mathcal{L}_{\mathbf{x}}$ as in Eq. (7), $\mathcal{L}_{\mathbf{p}}$ as in Eq. (8), and $\mathcal{L}_T$ defined as

$$\mathcal{L}_T = \sum_{i=1}^{3N}\left[\frac{\eta_1^{(i)}}{Q_1}\frac{\partial}{\partial p_i}\left(p_i\,\cdot\right) + \sum_{j=1}^{M_{\mathrm{NHC}}-1}\frac{p_{\eta_{j+1}^{(i)}}}{Q_{j+1}}\frac{\partial}{\partial p_{\eta_j^{(i)}}}\left(p_{\eta_j^{(i)}}\,\cdot\right) - \sum_{j=1}^{M_{\mathrm{NHC}}}G_j^{(i)}\frac{\partial}{\partial p_{\eta_j^{(i)}}} - \sum_{j=1}^{M_{\mathrm{NHC}}}\frac{p_{\eta_j^{(i)}}}{Q_j}\frac{\partial}{\partial \eta_j^{(i)}}\right] , \tag{138}$$

with $G_j^{(i)}$ defined by

$$\left.\begin{aligned} G_1^{(i)} &= \frac{p_i^2}{m_i} - k_B T \\ G_j^{(i)} &= \frac{p_{\eta_{j-1}^{(i)}}^2}{Q_{j-1}} - k_B T \quad \left(j = \overline{2, M_{\mathrm{NHC}}}\right) \end{aligned}\right\} \left(i = \overline{1, 3N}\right) . \tag{139}$$

The relevant phase space propagator $e^{\mathcal{L}_T \Delta t}$ for the $\mathcal{L}_T$ part for NHC may not be *exactly* obtained, because it involves nonlinear differential equations that are difficult to solve analytically. (For comparison, the exact expression for $e^{\mathcal{L}_T \Delta t}$ for a finite time interval $\Delta t$ in such as the Andersen or Langevin thermostat may be analytically derived such that Eq. (34) is satisfied.) Nevertheless, the numerical implementation of the phase space propagator $e^{\mathcal{L}_T \Delta t}$ for the NHC thermostat part may often be effectively accurate. The multiple time-scale technique such as the reference system propagator algorithm[13] (RESPA) and a higher-order (than $\Delta t^2$) factorization such as the Suzuki-Yoshida decomposition framework[28-30] may be used to guarantee the accuracy[27]. For instance, the equations of motion for the $\mathcal{L}_T$ part of NHC for a finite time interval $\Delta t$ may be expressed as[27]

$$
\left.\begin{array}{l}
p_{\eta^{(i)}_{M_{\mathrm{NHC}}}} \leftarrow p_{\eta^{(i)}_{M_{\mathrm{NHC}}}} + G^{(i)}_{M_{\mathrm{NHC}}} \frac{\delta_\alpha}{2} \\
\left.\begin{array}{l}
p_{\eta^{(i)}_j} \leftarrow p_{\eta^{(i)}_j} \exp\left(-\frac{p_{\eta^{(i)}_{j+1}}}{Q_{j+1}} \frac{\delta_\alpha}{4}\right) \\
p_{\eta^{(i)}_j} \leftarrow p_{\eta^{(i)}_j} + G^{(i)}_j \frac{\delta_\alpha}{2} \\
p_{\eta^{(i)}_j} \leftarrow p_{\eta^{(i)}_j} \exp\left(-\frac{p_{\eta^{(i)}_{j+1}}}{Q_{j+1}} \frac{\delta_\alpha}{4}\right)
\end{array}\right\} \left(j = \overline{M_{\mathrm{NHC}}-1, 1}\right) \\
\eta^{(i)}_j \leftarrow \eta^{(i)}_j + \frac{p_{\eta^{(i)}_j}}{Q_j} \delta_\alpha \qquad \left(j = \overline{1, M_{\mathrm{NHC}}}\right) \\
p_i \leftarrow p_i \exp\left(-\frac{p_{\eta^{(i)}_1}}{Q_1} \delta_\alpha\right) \\
\left.\begin{array}{l}
p_{\eta^{(i)}_j} \leftarrow p_{\eta^{(i)}_j} \exp\left(-\frac{p_{\eta^{(i)}_{j+1}}}{Q_{j+1}} \frac{\delta_\alpha}{4}\right) \\
p_{\eta^{(i)}_j} \leftarrow p_{\eta^{(i)}_j} + G^{(i)}_j \frac{\delta_\alpha}{2} \\
p_{\eta^{(i)}_j} \leftarrow p_{\eta^{(i)}_j} \exp\left(-\frac{p_{\eta^{(i)}_{j+1}}}{Q_{j+1}} \frac{\delta_\alpha}{4}\right)
\end{array}\right\} \left(j = \overline{1, M_{\mathrm{NHC}}-1}\right) \\
p_{\eta^{(i)}_{M_{\mathrm{NHC}}}} \leftarrow p_{\eta^{(i)}_{M_{\mathrm{NHC}}}} + G^{(i)}_{M_{\mathrm{NHC}}} \frac{\delta_\alpha}{2}
\end{array}\right\}
\begin{array}{l}
\left(\alpha = \overline{1, n_{\mathrm{SY}}}\right) \\
\left(k = \overline{1, n_{\mathrm{RESPA}}}\right) \\
\left(i = \overline{1, 3N}\right)
\end{array} . \qquad (140)
$$

Here, we use RESPA to divide an integration step for the NHC thermostat into $n_{\mathrm{RESPA}}$ equal parts, and implement the Suzuki-Yoshida decomposition framework[28-30] to further divide each part into $n_{\mathrm{SY}}$ smaller parts with different weights $\{w_\alpha\}$. The value of $n_{\mathrm{SY}}$ depends on the order of the Suzuki-Yoshida decomposition. Throughout our work the sixth order Suzuki-Yoshida factorization is employed. In this case, $n_{\mathrm{SY}} = 7$ and

$$
\begin{aligned}
w_1 &= w_7 = 0.784513610477560 \\
w_2 &= w_6 = 0.235573213359357 \\
w_3 &= w_5 = -1.17767998417887 \\
w_4 &= 1 - w_1 - w_2 - w_3 - w_5 - w_6 - w_7
\end{aligned}
\quad . \tag{141}
$$

The parameter $\delta_\alpha = \frac{w_\alpha}{n_{\mathrm{RESPA}}} \Delta t$ is the time step size for the $\alpha$-th of the $n_{\mathrm{SY}}$ smaller parts. When half a time interval $\Delta t/2$ is used for the physical degrees of freedom, the parameter becomes $\delta_\alpha = \frac{w_\alpha}{n_{\mathrm{RESPA}}} \frac{\Delta t}{2}$.

The conventional algorithm for NHC[12, 14, 27] employs the 'side' scheme [Eq. (27)]. The 'side-NHC' algorithm for propagating the MD trajectory through a time interval $\Delta t$ is

$$
\text{Eq. (140) with } \delta_\alpha = \frac{w_\alpha}{n_{\mathrm{RESPA}}} \frac{\Delta t}{2}, \tag{142}
$$

$$
\begin{aligned}
\mathbf{p} &\leftarrow \mathbf{p} - U'(\mathbf{x}) \frac{\Delta t}{2} \\
\mathbf{x} &\leftarrow \mathbf{x} + \mathbf{M}^{-1} \mathbf{p} \Delta t \\
\mathbf{p} &\leftarrow \mathbf{p} - U'(\mathbf{x}) \frac{\Delta t}{2}
\end{aligned}
\quad , \tag{143}
$$

$$
\text{Eq. (140) with } \delta_\alpha = \frac{w_\alpha}{n_{\mathrm{RESPA}}} \frac{\Delta t}{2} \ . \tag{144}
$$

Here, Eq. (142) and Eq. (144) share the same form as Eq. (140) except that the time step size for each smaller part is $\delta_\alpha = \frac{w_\alpha}{n_{\mathrm{RESPA}}} \frac{\Delta t}{2}$.

When the 'end' scheme [Eq. (28)] is used, the 'end-NHC' algorithm for propagating the MD trajectory through a time interval $\Delta t$ reads

$$\begin{aligned} &\mathbf{p} \leftarrow \mathbf{p} - U'(\mathbf{x})\frac{\Delta t}{2} \\ &\mathbf{x} \leftarrow \mathbf{x} + \mathbf{M}^{-1}\mathbf{p}\Delta t \quad , \\ &\mathbf{p} \leftarrow \mathbf{p} - U'(\mathbf{x})\frac{\Delta t}{2} \end{aligned} \tag{145}$$

$$\text{Eq. (140) with } \delta_\alpha = \frac{w_\alpha}{n_{\text{RESPA}}}\Delta t \ . \tag{146}$$

Similarly, when the 'middle' scheme [Eq. (29)] is implemented, the 'middle-NHC' algorithm for propagating the MD trajectory through a time interval $\Delta t$ is then

$$\begin{aligned} &\mathbf{p} \leftarrow \mathbf{p} - U'(\mathbf{x})\frac{\Delta t}{2} \\ &\mathbf{x} \leftarrow \mathbf{x} + \mathbf{M}^{-1}\mathbf{p}\frac{\Delta t}{2} \end{aligned} \quad , \tag{147}$$

$$\text{Eq. (140) with } \delta_\alpha = \frac{w_\alpha}{n_{\text{RESPA}}}\Delta t \ , \tag{148}$$

$$\begin{aligned} &\mathbf{x} \leftarrow \mathbf{x} + \mathbf{M}^{-1}\mathbf{p}\frac{\Delta t}{2} \\ &\mathbf{p} \leftarrow \mathbf{p} - U'(\mathbf{x})\frac{\Delta t}{2} \end{aligned} \quad . \tag{149}$$

When $M_{\text{NHC}} = 1$, NHC is reduced to the conventional Nosé-Hoover thermostat, which is easier to implement but more likely suffers the nonergodic problem[12]. It is trivial to obtain the Nosé-Hoover algorithms for the three schemes. Some similar work was done for the Nosé-Hoover thermostat by Itoh *et al*.[65]

## Appendix D. Thermostat integrators for path integral molecular dynamics

Imaginary time path integral maps a quantum system onto a classical ring polymer of 'beads' (i.e., replicas of the system) connected by harmonic springs[37, 66, 67]. Because fictitious momenta could be assigned to the beads, MD can then be employed to sample the path integral beads[38]. This approach is noted path integral molecular dynamics (PIMD), which offers a convenient and effective way for sampling quantum statistical properties in complex realistic systems[24, 26, 45, 46, 68, 69]. Quantum statistical effects (such as zero point energy, tunneling, etc.) become important at low temperatures and/or in molecular systems that contain light atoms (e.g., hydrogen, or helium).

### 1. Thermodynamic properties

Any thermodynamic property of the canonical ensemble is of the general form

$$\left\langle \hat{B} \right\rangle = \frac{1}{Z} \mathrm{Tr}\left( e^{-\beta \hat{H}} \hat{B} \right) , \tag{150}$$

where $Z = \mathrm{Tr}\left[ e^{-\beta \hat{H}} \right]$ is the partition function and $\hat{B}$ is an operator relevant to the specific property of interest. Eq. (150) can be expressed in the coordinate space $\mathbf{x}$, i.e.,

$$\left\langle \hat{B} \right\rangle = \frac{\int d\mathbf{x} \left\langle \mathbf{x} \right| e^{-\beta \hat{H}} \hat{B} \left| \mathbf{x} \right\rangle}{\int d\mathbf{x} \left\langle \mathbf{x} \right| e^{-\beta \hat{H}} \left| \mathbf{x} \right\rangle} \quad . \tag{151}$$

The denominator leads to

$$
\begin{aligned}
Z &= \int d\mathbf{x} \langle \mathbf{x} | e^{-\beta \hat{H}} | \mathbf{x} \rangle \\
&\overset{\mathbf{x}_1 \equiv \mathbf{x}}{=} \lim_{P\to\infty} \int d\mathbf{x}_1 \int d\mathbf{x}_2 \cdots \int d\mathbf{x}_P \left( \frac{P}{2\pi\beta\hbar^2} \right)^{3NP/2} |\mathbf{M}|^{P/2} \\
&\quad \times \exp\left\{ -\frac{P}{2\beta\hbar^2} \sum_{i=1}^{P} \left[ \left(\mathbf{x}_{i+1} - \mathbf{x}_i\right)^T \mathbf{M} \left(\mathbf{x}_{i+1} - \mathbf{x}_i\right) \right] - \frac{\beta}{P} \sum_{i=1}^{P} U\left(\mathbf{x}_i\right) \right\}
\end{aligned} \quad , \qquad (152)
$$

where $\mathbf{x}_{P+1} \equiv \mathbf{x}_1$ and $P$ is the number of path integral beads. Similarly, the numerator of Eq. (151) is

$$
\begin{aligned}
\int d\mathbf{x} \langle \mathbf{x} | e^{-\beta \hat{H}} \hat{B} | \mathbf{x} \rangle &\overset{\mathbf{x}_1 \equiv \mathbf{x}}{=} \lim_{P\to\infty} \int d\mathbf{x}_1 \int d\mathbf{x}_2 \cdots \int d\mathbf{x}_P \left( \frac{P}{2\pi\beta\hbar^2} \right)^{3NP/2} |\mathbf{M}|^{P/2} \\
&\quad \times \exp\left\{ -\frac{P}{2\beta\hbar^2} \sum_{i=1}^{P} \left[ \left(\mathbf{x}_{i+1} - \mathbf{x}_i\right)^T \mathbf{M} \left(\mathbf{x}_{i+1} - \mathbf{x}_i\right) \right] - \frac{\beta}{P} \sum_{i=1}^{P} U\left(\mathbf{x}_i\right) \right\} \\
&\quad \times \tilde{B}\left(\mathbf{x}_1, \cdots, \mathbf{x}_P\right)
\end{aligned} \quad . \qquad (153)
$$

It is straightforward to show that the estimator $\tilde{B}\left(\mathbf{x}_1, \cdots, \mathbf{x}_P\right)$ for any coordinate dependent operator $\hat{B}\left(\hat{\mathbf{x}}\right)$ is

$$
\tilde{B}\left(\mathbf{x}_1, \cdots, \mathbf{x}_P\right) = \frac{1}{P} \sum_{j=1}^{P} B\left(\mathbf{x}_j\right) \quad . \qquad (154)
$$

When $\hat{B} = \frac{1}{2} \hat{\mathbf{p}}^T \mathbf{M}^{-1} \hat{\mathbf{p}}$ is the kinetic energy operator, the primitive estimator is

$$
\tilde{B}\left(\mathbf{x}_1, \cdots, \mathbf{x}_P\right) = \frac{NP}{2\beta} - \sum_{j=1}^{P} \frac{P}{2\beta^2\hbar^2} \left[ \left(\mathbf{x}_{j+1} - \mathbf{x}_j\right)^T \mathbf{M} \left(\mathbf{x}_{j+1} - \mathbf{x}_j\right) \right] \qquad (155)
$$

and the virial version is

$$\tilde{B}(\mathbf{x}_1,\cdots,\mathbf{x}_P) = \frac{N}{2\beta} + \frac{1}{2P}\sum_{j=1}^{P}\left[\left(\mathbf{x}_j - \mathbf{x}^*\right)^T \frac{\partial U(\mathbf{x}_j)}{\partial \mathbf{x}_j}\right] \quad , \tag{156}$$

where $\mathbf{x}^*$ can be the centroid of the path integral beads [52]

$$\mathbf{x}^* = \mathbf{x}_c \equiv \frac{1}{P}\sum_{j=1}^{P}\mathbf{x}_j \tag{157}$$

or $\mathbf{x}^*$ can be any one of the *P* beads

$$\mathbf{x}^* = \mathbf{x}_i \quad , \tag{158}$$

with *i* fixed in Eq. (156).

**2. Staging Path Integral Molecular Dynamics**

Consider the staging transformation of Tuckerman *et al.* [27, 45, 46, 51]

$$\begin{aligned} \boldsymbol{\xi}_1 &= \mathbf{x}_1 \\ \boldsymbol{\xi}_j &= \mathbf{x}_j - \frac{(j-1)\mathbf{x}_{j+1} + \mathbf{x}_1}{j} \qquad \left(j = \overline{2,P}\right) \end{aligned} \; . \tag{159}$$

Its inverse transformation takes the recursive form

$$\begin{aligned} \mathbf{x}_1 &= \boldsymbol{\xi}_1 \\ \mathbf{x}_j &= \boldsymbol{\xi}_j + \frac{j-1}{j}\mathbf{x}_{j+1} + \frac{1}{j}\boldsymbol{\xi}_1 \qquad \left(j = \overline{2,P}\right) \end{aligned} \; . \tag{160}$$

Define

$$\omega_P = \frac{\sqrt{P}}{\beta\hbar} \ . \tag{161}$$

Eq. (152) becomes

$$\begin{aligned} Z &\overset{\xi_1 \equiv \mathbf{x}_1}{=} \lim_{P\to\infty} \left( \frac{P}{2\pi\beta\hbar^2} \right)^{3NP/2} |\mathbf{M}|^{P/2} \int d\xi_1 \int d\xi_2 \cdots \int d\xi_P \\ &\times \exp\left\{ -\beta \sum_{j=1}^{P} \left[ \frac{1}{2} \omega_P^2 \xi_j^T \bar{\mathbf{M}}_j \xi_j + \frac{1}{P} U\left( \mathbf{x}_j\left( \xi_1, \cdots, \xi_P \right) \right) \right] \right\} \end{aligned} , \tag{162}$$

with the (diagonal) mass matrices given by

$$\begin{aligned} &\bar{\mathbf{M}}_1 = 0 \\ &\bar{\mathbf{M}}_j = \frac{j}{j-1} \mathbf{M} \qquad \left( j = \overline{2, P} \right) \end{aligned} . \tag{163}$$

Define

$$\phi\left( \xi_1, \cdots, \xi_P \right) = \frac{1}{P} \sum_{j=1}^{P} U\left( \mathbf{x}_j\left( \xi_1, \cdots, \xi_P \right) \right) \ . \tag{164}$$

It is easy to verify the chain rule

$$\begin{aligned} &\frac{\partial \phi}{\partial \xi_1} = \sum_{i=1}^{P} \frac{\partial \phi}{\partial \mathbf{x}_i} = \frac{1}{P} \sum_{i=1}^{P} U'\left( \mathbf{x}_i \right) \\ &\frac{\partial \phi}{\partial \xi_j} = \frac{\partial \phi}{\partial \mathbf{x}_j} + \frac{j-2}{j-1} \frac{\partial \phi}{\partial \xi_{j-1}} \qquad \left( j = \overline{2, P} \right) \end{aligned} \tag{165}$$

from Eqs. (159)-(160). Employing the isomorphism strategy proposed by Chandler and Wolynes[37], one can insert fictitious momenta $\left( \mathbf{p}_1, \cdots, \mathbf{p}_P \right)$ into Eq. (162), which leads to

$$Z \overset{\xi_1 \equiv \mathbf{x}_1}{=} \lim_{P\to\infty} \left( \frac{P}{4\pi^2 \hbar^2} \right)^{3NP/2} |\mathbf{M}|^{P/2} \left( \prod_{j=1}^{P} |\tilde{\mathbf{M}}_j| \right)^{-1/2} \int \left( \prod_{j=1}^{P} d\xi_j d\mathbf{p}_j \right) \times \exp\left[ -\beta H_{eff} \left( \xi_1, \cdots, \xi_P ; \mathbf{p}_1, \cdots, \mathbf{p}_P \right) \right] \tag{166}$$

with the effective Hamiltonian given by

$$H_{\mathrm{eff}} \left( \xi_1, \cdots, \xi_P ; \mathbf{p}_1, \cdots, \mathbf{p}_P \right) = \sum_{j=1}^{P} \frac{1}{2} \mathbf{p}_j^T \tilde{\mathbf{M}}_j^{-1} \mathbf{p}_j + U_{\mathrm{eff}} \left( \xi_1, \cdots, \xi_P \right) \quad , \tag{167}$$

where

$$U_{\mathrm{eff}} \left( \xi_1, \cdots, \xi_P \right) = \sum_{j=1}^{P} \frac{1}{2} \omega_{\mathrm{P}}^2 \xi_j^T \bar{\mathbf{M}}_{\mathrm{j}} \xi_j + \phi\left( \xi_1, \cdots, \xi_P \right) \quad . \tag{168}$$

The fictitious masses are chosen as

$$\begin{aligned} \tilde{\mathbf{M}}_1 &= \mathbf{M} \\ \tilde{\mathbf{M}}_j &= \bar{\mathbf{M}}_j \qquad \left( j = \overline{2, P} \right) \end{aligned} \tag{169}$$

such that all staging modes $\left( \xi_2, \cdots, \xi_P \right)$ will move on the same time scale. The thermodynamic property Eq. (151) is then expressed as

$$\left\langle \hat{B} \right\rangle = \lim_{P\to\infty} \frac{\int \left( \prod_{j=1}^{P} d\xi_j d\mathbf{p}_j \right) \exp\left\{ -\beta H_{\mathrm{eff}} \left( \xi_1, \cdots, \xi_P ; \mathbf{p}_1, \cdots, \mathbf{p}_P \right) \right\} \tilde{B}\left( \mathbf{x}_1, \cdots, \mathbf{x}_P \right)}{\int \left( \prod_{j=1}^{P} d\xi_j d\mathbf{p}_j \right) \exp\left\{ -\beta H_{\mathrm{eff}} \left( \xi_1, \cdots, \xi_P ; \mathbf{p}_1, \cdots, \mathbf{p}_P \right) \right\}} \quad . \tag{170}$$

One may sample $\left( \xi_1, \cdots, \xi_P, \mathbf{p}_1, \cdots, \mathbf{p}_P \right)$ in a molecular dynamics (MD) scheme for evaluating the thermodynamic property. That is, Eq. (170) leads to

$$\begin{aligned} \dot{\boldsymbol{\xi}}_j &= \tilde{\mathbf{M}}_j^{-1}\,\mathbf{p}_j \\ \dot{\mathbf{p}}_j &= -\omega_P^2 \bar{\mathbf{M}}_j \boldsymbol{\xi}_j - \frac{\partial \phi}{\partial \boldsymbol{\xi}_j} \qquad \left( j = \overline{1,P} \right) \end{aligned} \quad . \tag{171}$$

The equations of motion for $\left(\boldsymbol{\xi}_1,\cdots,\boldsymbol{\xi}_P,\mathbf{p}_1,\cdots,\mathbf{p}_P\right)$ in Eq. (171) must be coupled to a thermostat to ensure a proper canonical distribution for $\left(\boldsymbol{\xi}_1,\cdots,\boldsymbol{\xi}_P,\mathbf{p}_1,\cdots,\mathbf{p}_P\right)$. Note that only the configurational distribution of PIMD is important in Eq. (170) for evaluating thermodynamic properties.

It is often claimed in conventional PIMD algorithms that it is more favorable to employ the decomposition of Eq. (171)

$$\begin{pmatrix} \dot{\boldsymbol{\xi}}_j \\ \dot{\mathbf{p}}_j \end{pmatrix} = \underbrace{\begin{pmatrix} \tilde{\mathbf{M}}_j^{-1}\,\mathbf{p}_j \\ -\omega_P^2 \bar{\mathbf{M}}_j \boldsymbol{\xi}_j \end{pmatrix}} + \underbrace{\begin{pmatrix} 0 \\ -\dfrac{\partial \phi}{\partial \boldsymbol{\xi}_j} \end{pmatrix}} \qquad \left( j = \overline{1,P} \right) \tag{172}$$

because the harmonic force term $-\omega_P^2 \bar{\mathbf{M}}_j \boldsymbol{\xi}_j$ often varies much more frequently than the force term $-\dfrac{\partial \phi}{\partial \boldsymbol{\xi}_j}$. Note that the exact solution to the first term of the RHS of Eq. (172) is available[26] (that is, the multiple time-scale technique such as RESPA[13] is applied). Our recent work[24], however, shows that

$$
\begin{pmatrix} \dot{\boldsymbol{\xi}}_j \\ \dot{\mathbf{p}}_j \end{pmatrix} = \underbrace{\begin{pmatrix} \tilde{\mathbf{M}}_j^{-1}\mathbf{p}_j \\ 0 \end{pmatrix}} + \underbrace{\begin{pmatrix} 0 \\ -\omega_P^2 \bar{\mathbf{M}}_j \boldsymbol{\xi}_j - \dfrac{\partial \phi}{\partial \boldsymbol{\xi}_j} \end{pmatrix}}
$$

$$
= \underbrace{\begin{pmatrix} \tilde{\mathbf{M}}_j^{-1}\mathbf{p}_j \\ 0 \end{pmatrix}} + \underbrace{\begin{pmatrix} 0 \\ -\dfrac{\partial U_{\text{eff}}}{\partial \boldsymbol{\xi}_j} \end{pmatrix}} \quad \left(j = \overline{1, P}\right) \tag{173}
$$

instead is a more accurate and more efficient decomposition when the 'middle' scheme is applied to the thermostat for PIMD.

When the Langevin thermostat is employed, it has been proved in Appendix C of Ref. [24] (and its Supplementary Material[70]) that Eq. (173) leads to the exact configurational distribution of the path integral beads in the harmonic limit, while Eq. (172) does not. It is trivial to show that the conclusion can be extended to any thermostat as long as the thermostat rigorously preserves the Maxwell momentum distribution in its thermostat step. For example, the Andersen thermostat has the same property when the 'middle' scheme is used for PIMD. When the NHC thermostat part is effectively accurate, it is expected that Eq. (173) is also numerically more favorable when NHC is used for thermostatting PIMD in the 'middle' scheme. (This is verified by the numerical results in Fig. 7 of the paper.)

Although the staging transformation of the path integral beads is used for demonstration, all conclusions hold for any other types of transformation of the beads (such as the normal mode transformation[26, 52, 70, 71]).

### 3. PIMD algorithms/integrators

The PIMD algorithms/integrators for such as the 'middle', 'side', and 'end' schemes share the same forms as their MD counterparts as listed in Appendix C. That is, replace the classical Hamiltonian [Eq. (1)], the phase space variables $(\mathbf{x},\mathbf{p})$, and the mass matrix $\mathbf{M}$ by the effective Hamiltonian [Eq. (167)], $(\xi_1,\cdots,\xi_P,\mathbf{p}_1,\cdots,\mathbf{p}_P)$, and $\{\tilde{\mathbf{M}}_j\}$, respectively. Similarly, the collision frequency $\nu$ is replaced by $\{\nu^{(l)}\ (l=\overline{1,P})\}$ for the $P$ staging coordinate variables $\{\xi_l\ (l=\overline{1,P})\}$ in the Andersen thermostat, the friction coefficient $\gamma$ by $\{\gamma^{(l)}\ (l=\overline{1,P})\}$ in Langevin dynamics, and the characteristic time $\tilde{\tau}_{\mathrm{NHC}}$ of Eq. (19) by $\{\tilde{\tau}_{\mathrm{NHC}}^{(l)}\ (l=\overline{1,P})\}$ in NHC. When staging PIMD is employed, the optimal values for friction coefficients $\{\gamma^{(l)}\ (l=\overline{2,P})\}$ are $\omega_P$ in the free particle limit, while those for collision frequencies $\{\nu^{(l)}\ (l=\overline{2,P})\}$ are $\sqrt{2}\omega_P$ in the free particle limit. It is trivial to extend the results to such as normal-mode PIMD.

## Appendix E. Thermostat parameters

Table 1 lists the parameters in the three thermostats used for MD simulations of the five systems in the paper.

The thermostat parameters of the first staging bead $\xi_1$ in staging PIMD are the same as those in MD for all thermostat (listed in Table 1). The Langevin friction coefficients for the rest $P$-1 staging beads $\xi_l\ (l=\overline{2,P})$ are all chosen to be $\gamma^{(l)}=\omega_P$, as suggested in our previous work[24]. Similarly, while the collision frequencies (in the Andersen thermostat) for the $P$-1 staging beads $\xi_l\ (l=\overline{2,P})$ are all chosen to be $\nu^{(l)}=\omega_P$ (around their optimal values $\nu^{(l)}=\sqrt{2}\omega_P$ in the free particle limit), the parameter $\tilde{\tau}_{\mathrm{NHC}}^{(l)}$ of Eq. (19) in NHC for those staging beads $\xi_l$ $\left(l=\overline{2,P}\right)$ are $\omega_P^{-1}$.

## Appendix F. Comparison between the velocity and position Verlet algorithms in the 'middle' scheme

When the thermostat vanishes, the schemes presented in the paper are reduced to the velocity Verlet algorithm for constant energy MD that generates the microcanonical ensemble. Alternatively, one may develop similar schemes using the position Verlet algorithm instead. For instance, when the position Verlet algorithm is employed instead of the velocity Verlet algorithm, the 'middle' scheme is then changed to

$$e^{\mathcal{L}\Delta t} \approx e^{\mathcal{L}^{\mathrm{Middle}}\Delta t} = e^{\mathcal{L}_{\mathbf{x}}\Delta t/2} e^{\mathcal{L}_{\mathbf{p}}\Delta t/2} e^{\mathcal{L}_T\Delta t} e^{\mathcal{L}_{\mathbf{p}}\Delta t/2} e^{\mathcal{L}_{\mathbf{x}}\Delta t/2} \ . \tag{174}$$

We note Eq. (174) the 'PV-middle' scheme. It is easy to verify that, when such as the Andersen thermostat or Langevin dynamics is employed, the stationary state distribution produced by the 'PV-middle' scheme for the harmonic system Eq. (30) is

$$\rho^{\text{PV-Middle}} = \frac{1}{\bar{Z}'_N} \exp\left[ -\beta \left( \frac{1}{2}\mathbf{p}^T (\mathbf{1} - \mathbf{M}^{-1}\mathbf{A}\frac{\Delta t^2}{4})\mathbf{M}^{-1}\mathbf{p} + \frac{1}{2}\left(\mathbf{x}-\mathbf{x}_{eq}\right)^T \mathbf{A}\left(\mathbf{x}-\mathbf{x}_{eq}\right) \right) \right] \ . \tag{175}$$

That is, the 'PV-middle' scheme also leads to the exact configurational distribution in the harmonic limit, regardless of any finite time interval $\Delta t$ (as long as the matrix $\mathbf{1}-\mathbf{M}^{-1}\mathbf{A}\frac{\Delta t^2}{4}$ is positive-definite). Fig. 8 compares the MD results for the average potential energy produced by the 'PV-middle' scheme to those given by the 'middle' scheme. It is demonstrated that the 'middle' scheme performs better in configurational sampling for anharmonic systems than the ‘PV-middle’ scheme does. The numerical performance of NHC is similar when the NHC thermostat part is effectively accurate, as supported by the results in Fig. 8.

In the harmonic limit, the ‘middle’ scheme always underestimates the average kinetic energy, while the ‘PV-middle’ scheme overestimates it. Similar behaviors are observed in the two anharmonic systems, as shown in Fig. 9 where the average kinetic energy is estimated by both schemes. In terms of accuracy as a function of the finite time interval $\Delta t$, the ‘middle’ scheme is also superior to the ‘PV-middle’ scheme in sampling the momentum space.

Similarly, the position Verlet algorithm can be implemented in other schemes (e.g., ‘side’ or ‘end’) to construct such as ‘PV-side’ or ‘PV-end’. None of these PV-type schemes performs better than the ‘middle’ scheme for configurational sampling, regardless of which type of thermostat is used.

## Tables and Figures

**Table 1.** Parameters for different thermostats for the five systems in the paper

| System | $\nu$ Andersen | $\gamma$ Langevin | $\tilde{\tau}_{\mathrm{NHC}}^{-1}$ NHC [(2)] |
|---|---|---|---|
| $U(x) = m\omega^2 x^2 / 2$ | 1.4 au | 1 au | 0.125 au |
| $U(x) = x^4 / 4$ | 1 au | 1 au | 0.125 au |
| $H_2O$ | 0.83 $fs^{-1}$ (0.02 au) [(1)] | 0.68 $fs^{-1}$ (0.0164 au) [(1)] | 0.083 $fs^{-1}$ (0.002 au) [(1)] |
| $(Ne)_{13}$ | 0.001 $fs^{-1}$ | 0.001 $fs^{-1}$ | 0.0008 $fs^{-1}$ |
| Liquid water | 0.005 $fs^{-1}$ | 0.005 $fs^{-1}$ | 0.00285 $fs^{-1}$ |

(1) $\hbar\nu$ or $\hbar\gamma$ is considered similar to $\hbar\omega$ when converting the parameters from atomic units to SI units. [E.g., see Eq. (116) or (119).]

(2) $M_{\mathrm{NHC}} = 4$ coupling thermostats in each chain; $n_{\mathrm{RESPA}} = 1, n_{\mathrm{SY}} = 7$ (the 6th order Suzuki-Yoshida factorization, its order of accuracy is $O(\Delta t^6)$) for $(Ne)_{13}$ and liquid water, $n_{\mathrm{RESPA}} = 4, n_{\mathrm{SY}} = 7$ for the one-dimensional harmonic and quartic model systems and the $H_2O$ molecule.

**Figure Captions**

**Fig. 1** (Color). MD results for the average potential energy using different time intervals. (a) The harmonic potential at $\beta = 8$. [Unit: atomic units (au)] (b) The quartic potential at $\beta = 8$. (Unit: au) (c) Average potential energy per atom $\langle U(x) \rangle / \left( N_{atom} k_B \right)$ (unit: Kelvin) for $H_2O$ at $T = 100$ K. The unit of the time interval is au in Panels (a)-(b), while that is femtosecond (fs) in Panel (c). Statistical error bars are included. The interval is increased until the propagation of the thermostat fails.

**Fig. 2** (Color). As in Fig. 1, but for MD results for the average kinetic energy using different time intervals. Exact value of kinetic energy for Panel (b) is 0.0625 au. Results of the 'end' scheme are closer to exact value than those of the 'side' scheme.

**Fig. 3** (Color). MD results for the average potential energy using different time intervals. (a) The harmonic potential at $\beta = 8$. (b) The quartic potential at $\beta = 8$. Atom units (au) are used. Statistical error bars are included. The time interval is increased until the propagation of the thermostat fails.

**Fig. 4** (Color). MD results for the averaged potential energy per atom $\langle U(\mathbf{x}) \rangle / \left( N_{\mathrm{atom}} k_{\mathrm{B}} \right)$ (unit: Kelvin) using different time intervals. (a) $H_2O$ at $T = 100$ K. (b) $(Ne)_{13}$ at $T = 14$ K. (c) Liquid water at $T = 298.15$ K. Statistical error bars are included.

**Fig. 5** (Color). MD results for the average kinetic energy using different time intervals. (a) The harmonic potential at $\beta = 8$. (b) The quartic potential at $\beta = 8$. The units of both the energy and the time interval are atomic units (au). Statistical error bars are included.

**Fig. 6** (Color). MD results for the averaged kinetic energy per atom $\langle \mathbf{p}^T \mathbf{M}^{-1} \mathbf{p} \rangle / (2N_{\text{atom}} k_B)$ (unit: Kelvin) using different time intervals. (a) $H_2O$ at $T = 100$ K. (b) $(Ne)_{13}$ at $T = 14$ K. (c) Liquid water at $T = 298.15$ K. The unit of the time interval is femtosecond (fs). Statistical error bars are included.

**Fig. 7** (Color). PIMD results using different time intervals for liquid water at $T = 298.15$ K. (a) The average kinetic energy per atom $\langle \mathbf{p}^T \mathbf{M}^{-1} \mathbf{p} \rangle / (2N_{\text{atom}} k_{\text{B}})$ (unit: Kelvin). The primitive estimator is used. (b) Absolute difference between the primitive and virial estimators (unit: Kelvin). (c) The averaged potential energy per atom $\langle U(\mathbf{x}) \rangle / (N_{\text{atom}} k_{\text{B}})$ (unit: Kelvin). Statistical error bars are included.

**Fig. 8** (Color). MD results for the average potential energy using different time intervals. (a) The harmonic potential at $\beta = 8$. (Unit: au) (b) The quartic potential at $\beta = 8$. (Unit: au) (c) potential energy per atom $\langle U(x) \rangle / (N_{\text{atom}} k_B)$ (Unit: Kelvin) for $H_2O$ at $T = 100$ K. The units are au in Panels (a) and (b), while that of the time interval is femtosecond (fs) in Panel (c). Statistical error bars are included. The interval is increased until the propagation of the thermostat fails.

**Fig. 9** (Color). As in Fig. 8, but for MD results for the average kinetic energy using different time intervals.

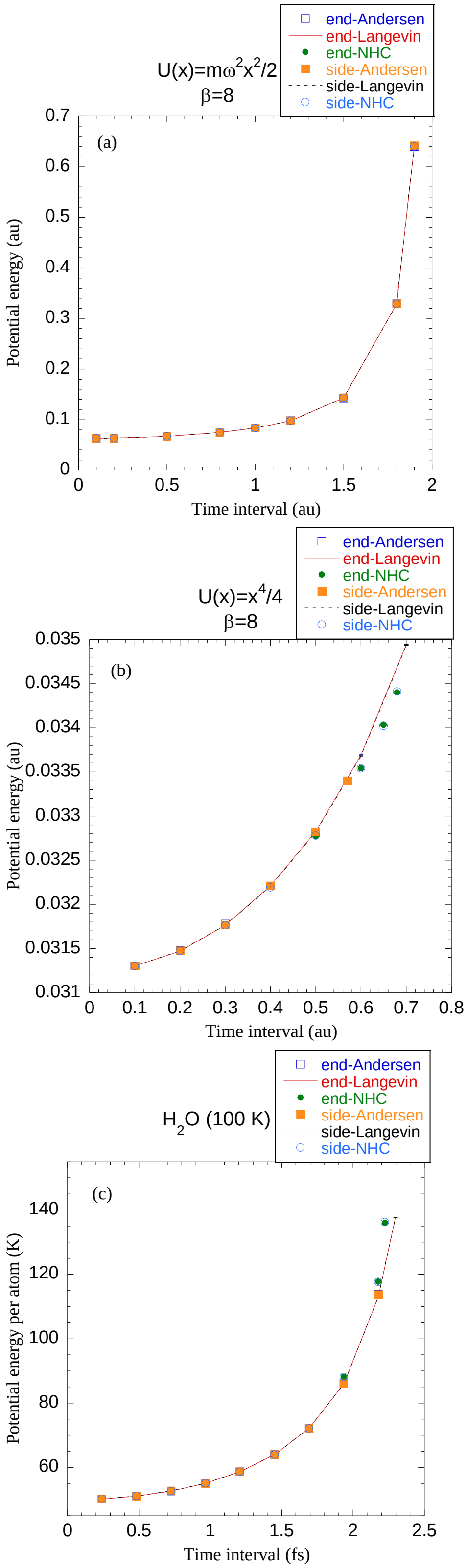

**Fig. 1**

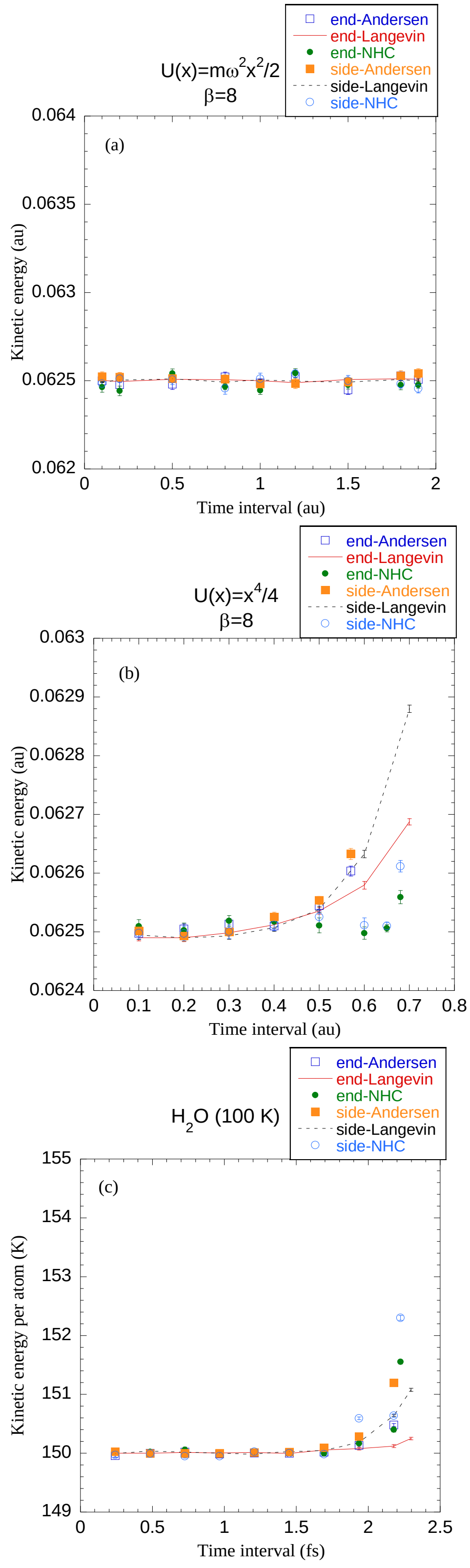

**Fig. 2**

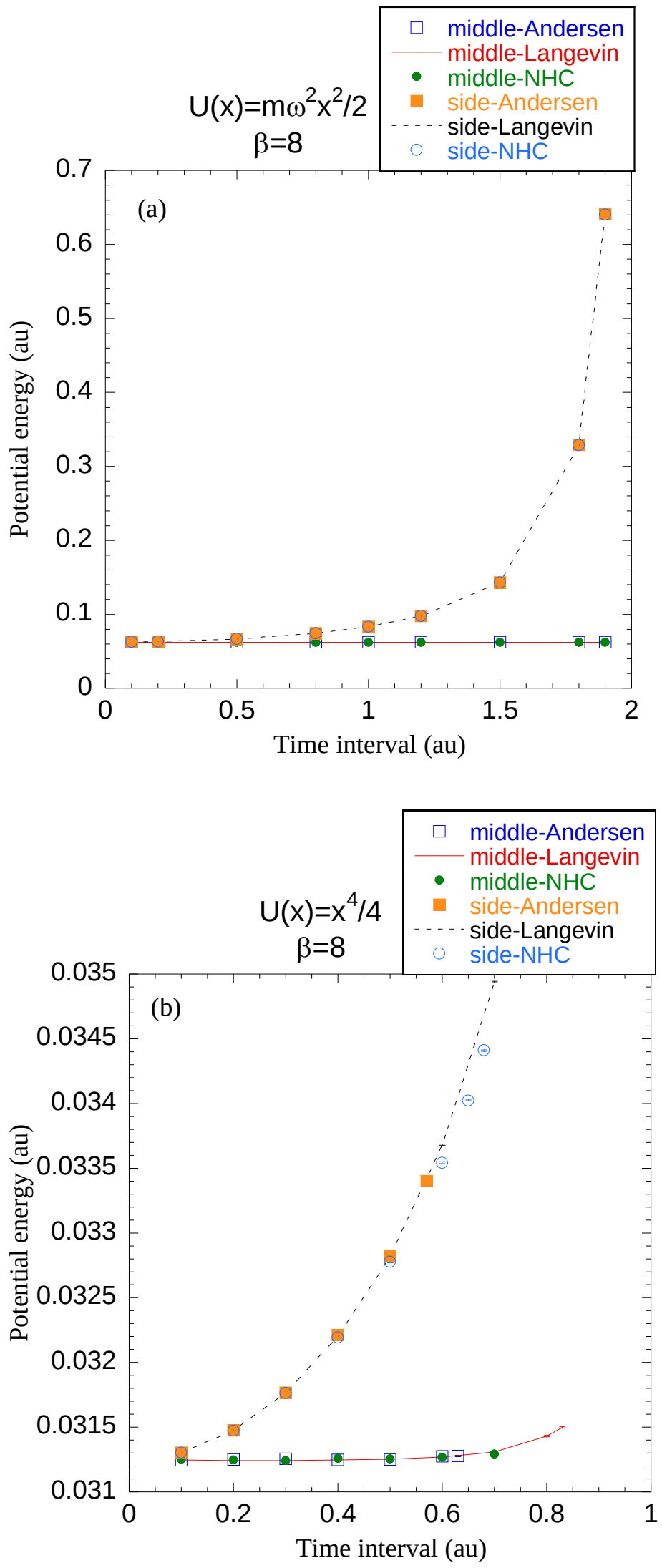

**Fig. 3**

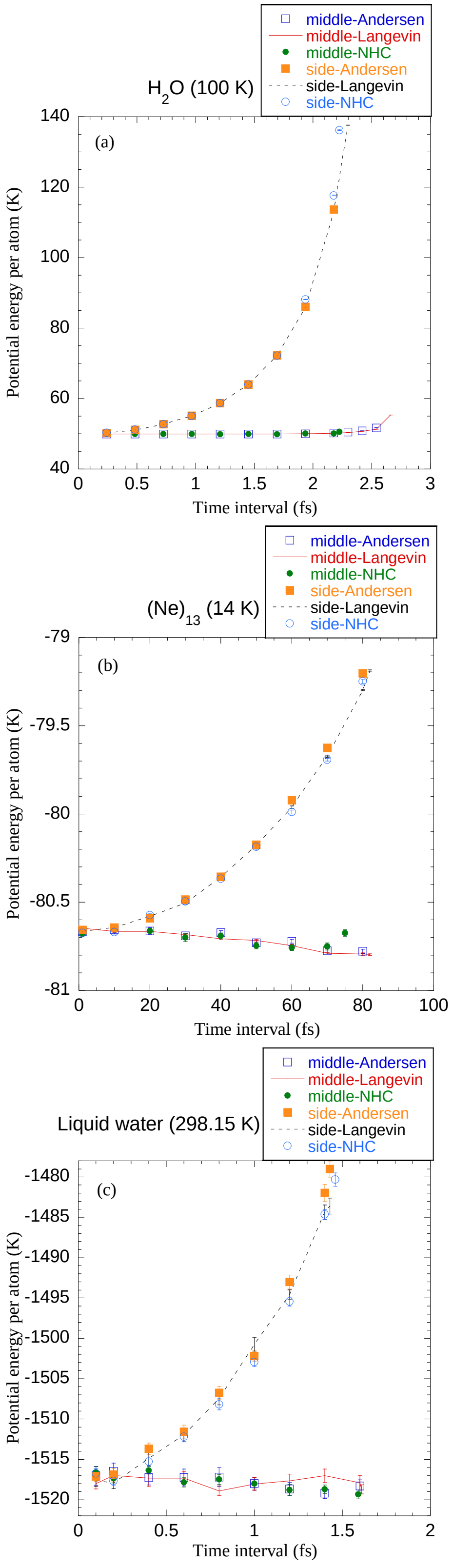

**Fig. 4**

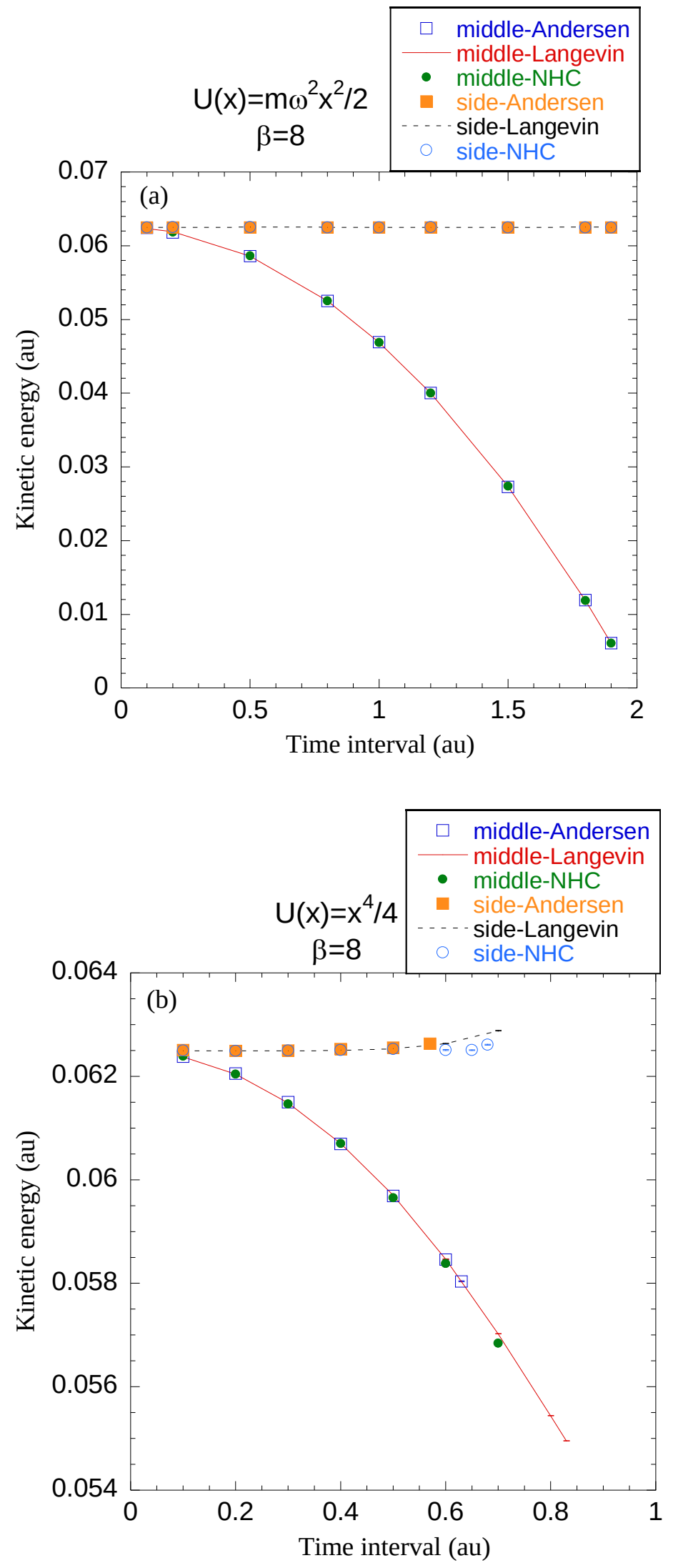

**Fig. 5**

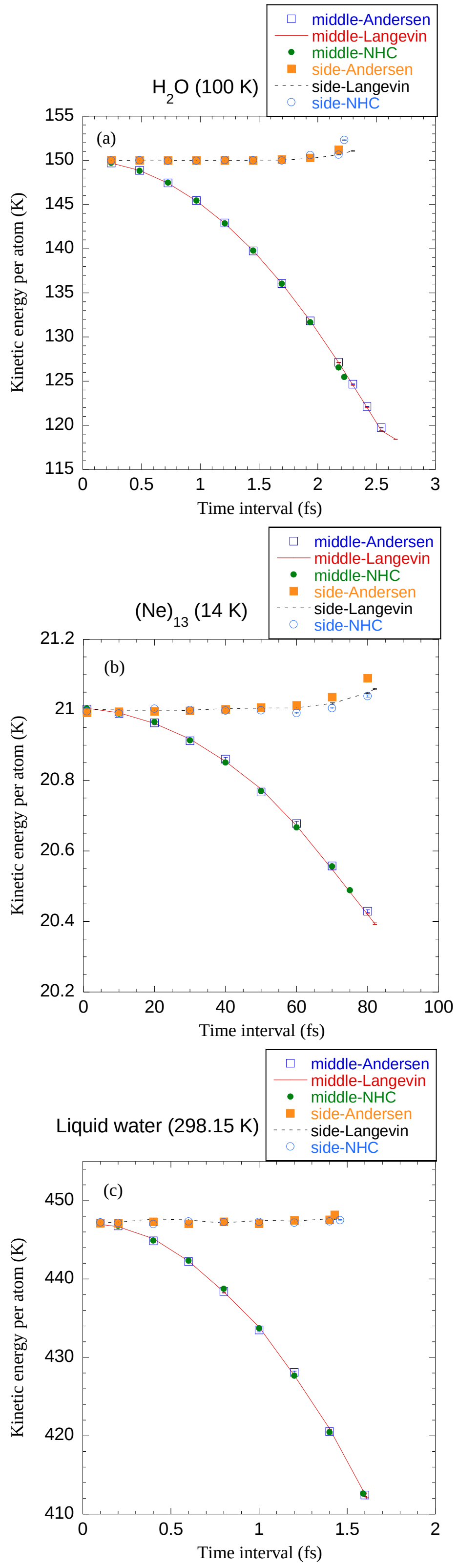

**Fig. 6**

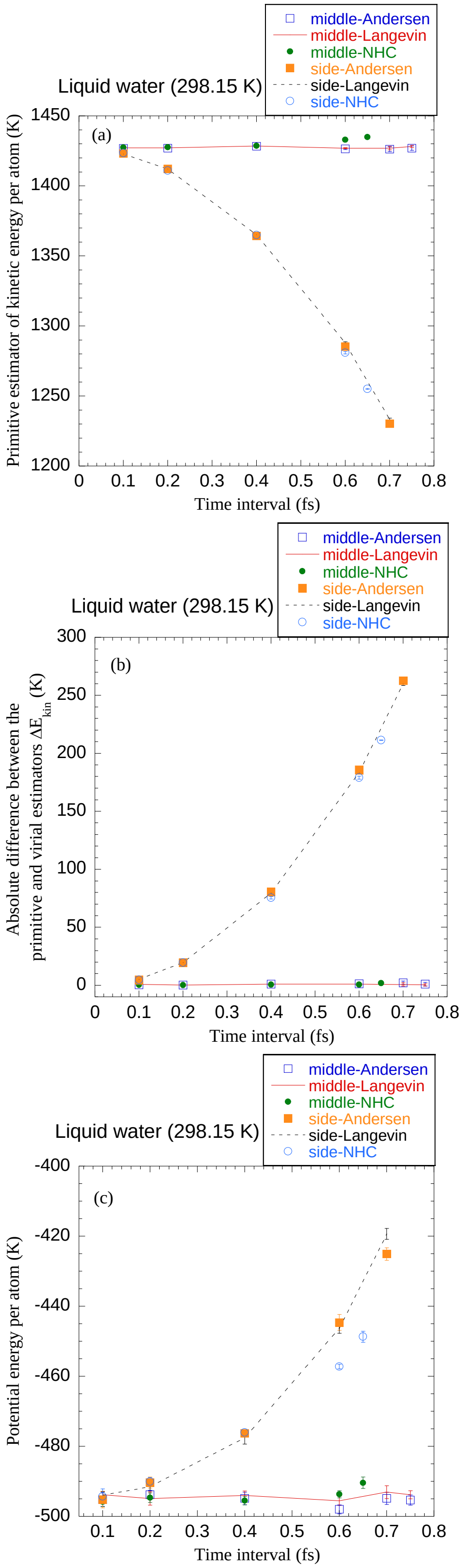

Liquid water (298.15 K)
middle-Andersen
middle-Langevin
middle-NHC
side-Andersen
side-Langevin
side-NHC
(a)
Primitive estimator of kinetic energy per atom (K)
Time interval (fs)
(b)
Absolute difference between the primitive and virial estimators ΔE_kin (K)
Time interval (fs)
(c)
Potential energy per atom (K)
Time interval (fs)

**Fig. 7**

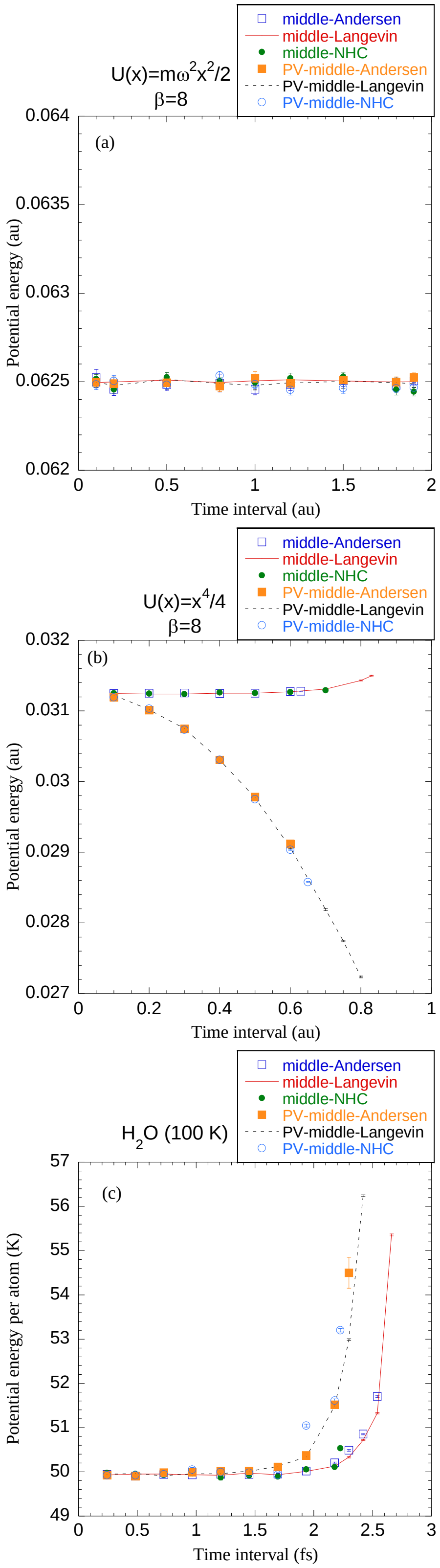

**Fig. 8**

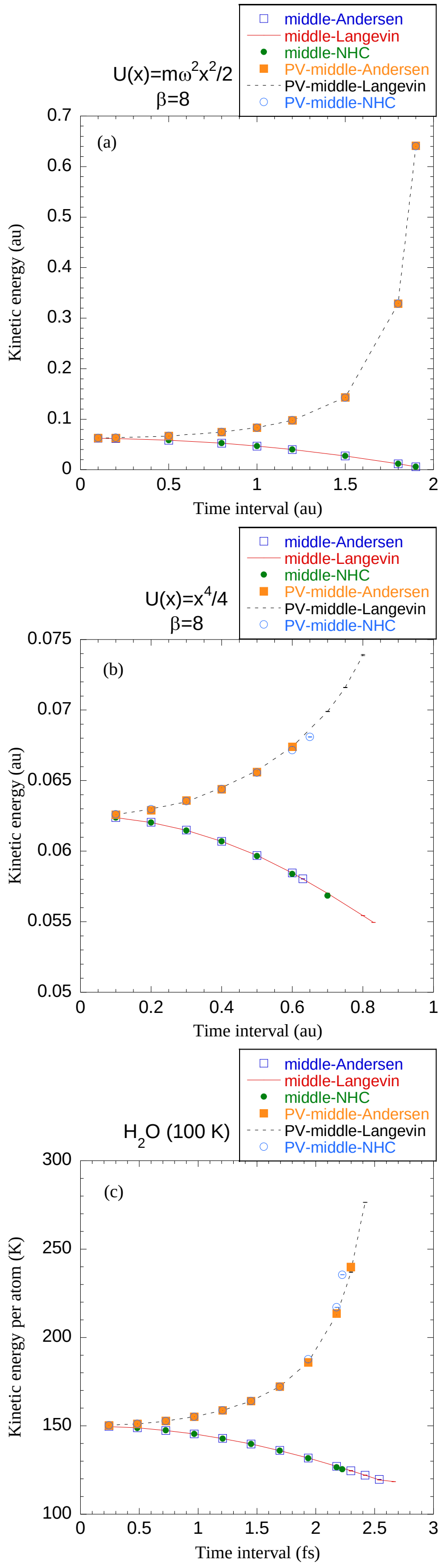

**Fig. 9**

# Supplementary material for
# “A unified thermostat scheme for efficient configurational sampling for classical/quantum canonical ensembles *via* molecular dynamics”

*Zhijun Zhang*[†], *Xinzijian Liu*[†], *Zifei Chen, Haifeng Zheng, Kangyu Yan, and Jian Liu**

Beijing National Laboratory for Molecular Sciences, Institute of Theoretical and Computational Chemistry, College of Chemistry and Molecular Engineering,

Peking University, Beijing 100871, China

* Electronic mail: jianliupku@pku.edu.cn

† Both authors contributed equally to the work.

**S1. Supplementary material for Appendix A on the Andersen thermostat**

Consider the Andersen thermostat for the harmonic system $U(x) = A(x - x_{\mathrm{eq}})^2 / 2$. Below we show the convergence of the $k$th-order moments.

Our aim is to prove that the spectral radius of the $(k+1)\times(k+1)$ matrix

$$\tilde{\mathbf{A}} \equiv \mathbf{A}_1\mathbf{A}_2\mathbf{A}_3\mathbf{A}_2\mathbf{A}_1 \tag{S1}$$

is less than 1, i.e. $SR(\tilde{\mathbf{A}}) < 1$. The spectral radius of a matrix is the largest absolute value of its eigenvalues, i.e. $SR(\tilde{\mathbf{A}}) = \max\left\{\left|\lambda_i(\tilde{\mathbf{A}})\right|\right\}$. In Eq. (S1)

$$\mathbf{A}_1 = \begin{pmatrix} 1 & & & \\ -A\frac{\Delta t}{2} & 1 & & \\ \vdots & \ddots & \ddots & \\ \left(-A\frac{\Delta t}{2}\right)^k & \cdots & -A\frac{k\Delta t}{2} & 1 \end{pmatrix} \quad \mathbf{A}_2 = \begin{pmatrix} 1 & \frac{k\Delta t}{2m} & \cdots & \left(\frac{\Delta t}{2m}\right)^k \\ & 1 & \ddots & \vdots \\ & & \ddots & \frac{\Delta t}{2m} \\ & & & 1 \end{pmatrix} . \tag{S2}$$

$$\mathbf{A}_3 = \begin{pmatrix} 1 & & & \\ & e^{-\nu\Delta t} & & \\ & & \ddots & \\ & & & e^{-\nu\Delta t} \end{pmatrix}$$

Here, parameters $A > 0,\ m > 0,\ \nu > 0$ and $0 < \Delta t < \sqrt{\frac{4m}{A}}$.

Consider a sequence of random variables $\{(x_n, p_n) : n = 1, 2, \cdots\}$ that satisfy

$$\begin{aligned} p_{n-1}' &\equiv p_{n-1} - Ax_{n-1}\frac{\Delta t}{2} \\ x_{n-1}' &\equiv x_{n-1} + \frac{p_{n-1}'}{m}\frac{\Delta t}{2} \\ p_{n-1}'' &\equiv \begin{cases} 0, & \text{with probability } 1 - e^{-\nu\Delta t} \\ p_{n-1}', & \text{with probability } e^{-\nu\Delta t} \end{cases} , \\ x_n &= x_{n-1}' + \frac{p_{n-1}''}{m}\frac{\Delta t}{2} \\ p_n &= p_{n-1}'' - Ax_n\frac{\Delta t}{2} \end{aligned} \tag{S3}$$

i.e. the 'middle-Andersen' algorithm for the harmonic system when the temperature is zero. Define the $k$th-order moment vector $\zeta_n^{(k)} = \left( \left\langle x_n^{k-j} p_n^j \right\rangle, j = 0, \cdots, k \right)^T$. It is trivial to show that the iteration of the moment vector satisfies

$$\zeta_n^{(k)} = \mathbf{A}_1 \mathbf{A}_2 \mathbf{A}_3 \mathbf{A}_2 \mathbf{A}_1 \zeta_{n-1}^{(k)} = \tilde{\mathbf{A}} \zeta_{n-1}^{(k)}, \quad n = 1, 2, \cdots \quad . \tag{S4}$$

Note that, matrix $\tilde{\mathbf{A}}$ for the zero-temperature case is the same as that for the finite temperature case. It is easy to show that $SR\left(\tilde{\mathbf{A}}\right) < 1$ if and only if

$$\zeta_n^{(k)} = \tilde{\mathbf{A}}^n \zeta_0^{(k)} \to 0, \quad n \to \infty, \text{ for all } \zeta_0^{(k)} \tag{S5}$$

by definition. Eq. (S5) can be derived from that $\left(x_n, p_n\right)$ converges to $\left(0,0\right)$ in probability and that $\left(x_n, p_n\right)$ is uniformly bounded. We first verify that $\left(x_n, p_n\right)$ converges to $\left(0,0\right)$ in probability, i.e., $\left(x_n, p_n\right) \xrightarrow{P} \left(0,0\right)$.

Expressing Eq. (S3) in the matrix form, we have

$$\begin{pmatrix} x_n \\ p_n \end{pmatrix} = \mathbf{B}_1 \mathbf{B}_2^2 \mathbf{B}_1 \begin{pmatrix} x_{n-1} \\ p_{n-1} \end{pmatrix} \tag{S6}$$

with probability $e^{-\nu \Delta t}$ and

$$\begin{pmatrix} x_n \\ p_n \end{pmatrix} = \mathbf{B}_1 \mathbf{B}_2 \begin{pmatrix} 1 & 0 \\ 0 & 0 \end{pmatrix} \mathbf{B}_2 \mathbf{B}_1 \begin{pmatrix} x_{n-1} \\ p_{n-1} \end{pmatrix} \tag{S7}$$

with probability $1 - e^{-\nu \Delta t}$. Here,

$$\mathbf{B}_1 = \begin{pmatrix} 1 & 0 \\ -A\dfrac{\Delta t}{2} & 1 \end{pmatrix} \quad \mathbf{B}_2 = \begin{pmatrix} 1 & \dfrac{\Delta t}{2m} \\ 0 & 1 \end{pmatrix} \tag{S8}$$

The general form of $\left(x_n, p_n\right)^T$ may then be expressed as

$$\begin{aligned} \begin{pmatrix} x_n \\ p_n \end{pmatrix} &= \mathbf{B}_1 \mathbf{B}_2 \mathbf{B}^{n - \sum_{i=0}^r n_i} \prod_{i=1}^r \left[ \begin{pmatrix} 1 & 0 \\ 0 & 0 \end{pmatrix} \mathbf{B}^{n_i} \right] \begin{pmatrix} 1 & 0 \\ 0 & 0 \end{pmatrix} \mathbf{B}^{n_0 - 1} \mathbf{B}_2 \mathbf{B}_1 \begin{pmatrix} x_0 \\ p_0 \end{pmatrix} \\ &= \mathbf{B}_1 \mathbf{B}_2 \mathbf{B}^{n - \sum_{i=0}^r n_i} \prod_{i=1}^r \left[ \begin{pmatrix} 1 & 0 \\ 0 & 0 \end{pmatrix} \mathbf{B}^{n_i} \begin{pmatrix} 1 & 0 \\ 0 & 0 \end{pmatrix} \right] \mathbf{B}^{n_0 - 1} \mathbf{B}_2 \mathbf{B}_1 \begin{pmatrix} x_0 \\ p_0 \end{pmatrix} \end{aligned}, \tag{S9}$$

where $\mathbf{B} \equiv \mathbf{B}_2 \mathbf{B}_1^2 \mathbf{B}_2$. In Eq. (S9) independent and identically distributed random variables $n_i$, $i = 0, 1, \cdots$ share the geometric distribution

$$P\left(n_i = j\right) = \left(e^{-\nu\Delta t}\right)^{j-1}\left(1 - e^{-\nu\Delta t}\right), \quad j = 1, 2, \cdots \quad . \tag{S10}$$

Let $r$ be an integer such that $\sum_{i=0}^{r} n_i \le n$ and $\sum_{i=0}^{r+1} n_i > n$. The eigen-decomposition of $\mathbf{B}$ gives $\mathbf{B} = \mathbf{P}\mathbf{\Lambda}\mathbf{P}^{-1}$ with

$$\mathbf{P} = \begin{pmatrix} 1 & 1 \\ \dfrac{2A\Delta t m^2}{\sqrt{A\Delta t^2 m^2\left(A\Delta t^2 - 4m\right)}} & -\dfrac{2A\Delta t m^2}{\sqrt{A\Delta t^2 m^2\left(A\Delta t^2 - 4m\right)}} \end{pmatrix} \tag{S11}$$

$$\mathbf{P}^{-1} = \begin{pmatrix} \dfrac{1}{2} & \dfrac{\Delta t\left(A\Delta t^2 - 4m\right)}{4\sqrt{A\Delta t^2 m^2\left(A\Delta t^2 - 4m\right)}} \\ \dfrac{1}{2} & -\dfrac{\Delta t\left(A\Delta t^2 - 4m\right)}{4\sqrt{A\Delta t^2 m^2\left(A\Delta t^2 - 4m\right)}} \end{pmatrix}, \tag{S12}$$

and

$$\begin{aligned} &\mathbf{\Lambda} = \mathrm{diag}\left\{\lambda_1, \lambda_2\right\} \\ &\lambda_{1,2} = 1 - \frac{A\Delta t^2}{2m} \mp i\sqrt{\frac{A\Delta t^2}{m}\left(1 - \frac{A\Delta t^2}{4m}\right)} \end{aligned} \quad . \tag{S13}$$

Eqs. (S11)-(S13) lead to

$$\begin{pmatrix} 1 & 0 \\ 0 & 0 \end{pmatrix}\mathbf{B}^j\begin{pmatrix} 1 & 0 \\ 0 & 0 \end{pmatrix} = \begin{pmatrix} \dfrac{\lambda_1^j + \lambda_2^j}{2} & 0 \\ 0 & 0 \end{pmatrix} \quad . \tag{S14}$$

Note $\left|\lambda_{1,2}\right| = 1$. We introduce parameter $\varphi = \arccos\left(1 - \dfrac{A\Delta t^2}{2m}\right)$ such that $\lambda_{1,2} = e^{\mp i\varphi}$ and

$$\begin{pmatrix} 1 & 0 \\ 0 & 0 \end{pmatrix}\mathbf{B}^j\begin{pmatrix} 1 & 0 \\ 0 & 0 \end{pmatrix} = \begin{pmatrix} \cos j\varphi & 0 \\ 0 & 0 \end{pmatrix} \quad . \tag{S15}$$

So Eq. (S9) becomes

$$\begin{pmatrix} x_n \\ p_n \end{pmatrix} = \mathbf{B}_1\mathbf{B}_2\mathbf{B}^{n - \sum_{i=0}^{r} n_i}\begin{pmatrix} \prod_{i=1}^{r}\cos n_i\varphi & 0 \\ 0 & 0 \end{pmatrix}\mathbf{B}^{n_0 - 1}\mathbf{B}_2\mathbf{B}_1\begin{pmatrix} x_0 \\ p_0 \end{pmatrix} \quad . \tag{S16}$$

Consider the term $\prod_{i=1}^{r}\cos n_i\varphi$. Denote $P(\ )$ the probability function. The relation

$$P\left(\left|\cos n_i\varphi\right| > \left|\cos\varphi\right|\right) \le P(n_i \ne 1) = 1 - P\left(n_i = 1\right), \quad i = 1,2,\cdots. \qquad \text{(S17)}$$

always holds. Using $\cos\varphi = 1 - \frac{A\Delta t^2}{2m}$ and Eq. (S10), one then obtains

$$P\left(\left|\cos n_i\varphi\right| > \left|1 - \frac{A\Delta t^2}{2m}\right|\right) \le 1 - P\left(n_i = 1\right) = e^{-\nu\Delta t}, \quad i = 1,2,\cdots \ . \qquad \text{(S18)}$$

Define independent and identically distributed random variables $X_i \equiv \left|\cos n_i\varphi\right|, \quad i = 1,2,\cdots$ . Denote $a \equiv \left|1 - \frac{A\Delta t^2}{2m}\right| < 1$ , $b \equiv e^{-\nu\Delta t} < 1$ , and $P_a \equiv P\left(X_i > a\right)$ such that Eq. (S18) is expressed as $P_a \le b$.

Divide the set $\left\{X_i : i = 1,2,\cdots,r\right\}$ into two subsets. All variables in subset $\left\{A_j : j = 1,\cdots,n_A\right\} \equiv \left\{X_i : 1 \le i \le r, X_i \le a\right\}$ are not larger than $a$, where $0 \le n_A \le r$, while all variables in the other subset $\left\{B_j : j = 1,\cdots,n_B\right\} \equiv \left\{X_i : 1 \le i \le r, X_i > a\right\}$ are larger than $a$, where $n_B = r - n_A$. Since $0 \le X_i \equiv \left|\cos n_i\varphi\right| \le 1$, the product of the elements in the set $\left\{X_i\right\}$ is not larger than the product of the elements in the subset $\left\{A_j\right\}$. Therefore

$$\prod_{i=1}^{r} X_i \le \prod_{j=1}^{n_A} A_j \le a^{n_A}. \qquad \text{(S19)}$$

For all $\varepsilon > 0$, the probability

$$\begin{aligned} P\left(\prod_{i=1}^{r} X_i > \varepsilon\right) &\le P\left(a^{n_A} > \varepsilon\right) = P\left(n_A < \log\varepsilon / \log a\right) \\ &\le P\left(n_A \le \left\lfloor \log\varepsilon / \log a \right\rfloor + 1\right) \end{aligned}, \qquad \text{(S20)}$$

where $\lfloor x \rfloor$ stands for the floor function of a real number $x$, which is the largest integer that is not larger than $x$ . Let $N_1 \equiv \left\lfloor \log\varepsilon / \log a \right\rfloor + 1$ . Recalling

$\{A_j\} \equiv \{X_i : 1 \le i \le r, X_i \le a\}$, we obtain

$$\begin{aligned} P(n_A \le N_1) &= \sum_{i=0}^{N_1} P(n_A = i) \\ &= \sum_{i=0}^{N_1} C_r^i P_a^{r-i} (1-P_a)^i \\ &\le \sum_{i=0}^{N_1} C_r^i P_a^{r-i} \end{aligned} \quad . \tag{S21}$$

because $X_1, X_2, \cdots$ are independent. Here $C_r^i = \frac{r!}{i!(r-i)!}$ is the binomial coefficient. Because $P_a \le b$, the following relation holds,

$$\begin{aligned} P(n_A \le N_1) &\le \sum_{i=0}^{N_1} \frac{r!}{i!(r-i)!} b^{r-i} \\ &\le (N_1+1) r^{N_1} b^{r-N_1} \to 0, \;\; r \to \infty \end{aligned} \quad . \tag{S22}$$

Therefore, for all $\delta > 0$, there exists an integer $N_2$ satisfying $(N_1+1) r^{N_1} b^{r-N_1} < \delta$ when $r > N_2$. Let $N = \max\{N_1, N_2\}$. For all $\varepsilon > 0$, $\delta > 0$, when $r > N$, $P\left(\prod_{i=1}^{r} X_i > \varepsilon\right) < \delta$, which means $\prod_{i=1}^{r} X_i \xrightarrow{P} 0$ by definition.

For all $\varepsilon > 0, \delta > 0$, there then exists an integer $N_3$ satisfying $P\left(\prod_{i=1}^{r} |\cos n_i \varphi| > \varepsilon\right) < \frac{\delta}{2}$ when $r > N_3$. Consider the case $r \le N_3$, due to $\sum_{i=0}^{r+1} n_i > n$, a necessary condition is that there exists an integer $i$, satisfying $0 \le i \le r+1$ and $n_i \ge \frac{n}{N_3+2}$. Thus

$$P(r \le N_3) \le \sum_{i=0}^{r+1} P\left(n_i \ge \frac{n}{N_3+2}\right) \le (N_3+2)\left(e^{-\nu\Delta t}\right)^{\frac{n}{N_3+2}-1} \quad . \tag{S23}$$

Let $N_4 = \left\lfloor (N_3+2)\left[1 + \frac{1}{-\nu\Delta t} \log \frac{\delta}{2(N_3+2)}\right] \right\rfloor + 1$. When $n > N_4$,

$$P(r \le N_3) < (N_3+2) \frac{\delta}{2(N_3+2)} = \frac{\delta}{2} \quad . \tag{S24}$$

Let $\bar{N} = \max\{N_3, N_4\}$. When $n > \bar{N}$, we have

$$P\left(\prod_{i=1}^{r}|\cos n_i\varphi| > \varepsilon\right) \le P\left(\prod_{i=1}^{r}|\cos n_i\varphi| > \varepsilon \,\middle|\, r > N_3\right) + P(r \le N_3) < \frac{\delta}{2} + \frac{\delta}{2} = \delta \quad . \quad \text{(S25)}$$

Here, $P\left(\prod_{i=1}^{r}|\cos n_i\varphi| > \varepsilon \,\middle|\, r > N_3\right)$ is the conditional probability for $\prod_{i=1}^{r}|\cos n_i\varphi| > \varepsilon$ under the condition $r > N_3$. That is, $\prod_{i=1}^{r}|\cos n_i\varphi| \xrightarrow{P} 0,\ n \to \infty$. The $L^2$ norm of $(x_n, p_n)^T$, denoted as $\left\|(x_n, p_n)^T\right\|_2$, is given by $\sqrt{x_n^2 + p_n^2}$. Consider a $2\times 2$ matrix $\tilde{\mathbf{B}}$. Its matrix norm induced by the $L^2$ vector norm is defined as $\|\tilde{\mathbf{B}}\|_2 = \sup\left\{\|\tilde{\mathbf{B}}\tilde{\mathbf{x}}\|_2 :\ \text{any } 2-\text{dimensional vector } \tilde{\mathbf{x}} \text{ with } \|\tilde{\mathbf{x}}\|_2 = 1\right\}$. ( $\sup\{\ \}$ represents the superior limit.) As $\|\mathbf{B}^j\|_2$ is uniformly bounded for all $j$, Eq. (S16) leads to

$$\left\|\begin{pmatrix} x_n \\ p_n \end{pmatrix}\right\|_2 \le \|\mathbf{B}_1\mathbf{B}_2\|_2 \left\|\mathbf{B}^{n-\sum_{i=0}^{r} n_i}\right\|_2 \left\|\begin{pmatrix} \prod_{i=1}^{r}\cos n_i\varphi & 0 \\ 0 & 0 \end{pmatrix}\right\|_2 \left\|\mathbf{B}^{n_0-1}\right\|_2 \|\mathbf{B}_2\mathbf{B}_1\|_2 \left\|\begin{pmatrix} x_0 \\ p_0 \end{pmatrix}\right\|_2 \xrightarrow{P} 0 \quad \text{(S26)}$$

which is equivalent to $(x_n, p_n) \xrightarrow{P} (0,0)$.

Due to the uniform boundedness of $\|\mathbf{B}^j\|_2$, there exists $D > 0$ satisfying $\|(x_n, p_n)\|_2 \le D\|(x_0, p_0)\|_2$ for all $n$. The convergence in probability is then equivalent to the convergence in the *k*-th mean, i.e. $\zeta_n^{(k)} = \tilde{\mathbf{A}}^n \zeta_0^{(k)} \to 0$ for all $\zeta_0^{(k)}$, which leads to the statement $SR(\tilde{\mathbf{A}}) < 1$.

## S2. The optimal values for the thermostat parameters for MD simulations

Although any non-zero collision frequency $\nu$ in the Andersen thermostat or any non-zero friction coefficient $\gamma$ in Langevin dynamics can in principle generate the canonical distribution, it takes too long for the system to reach the canonical equilibrium if $\nu$ or $\gamma$ is too small. When the collision frequency $\nu$ or the friction coefficient $\gamma$ becomes zero, the dynamics is reduced to constant energy MD that generates the microcanonical ensemble. On the other hand, when the collision frequency $\nu$ or the friction coefficient $\gamma$ is too large, an established equilibrium becomes unstable leading to large statistic errors.

One should also choose a reasonable value for the NHC time parameter $\tilde{\tau}_{\mathrm{NHC}}$ of Eq.(19). A rational choice for $\tilde{\tau}_{\mathrm{NHC}}$ should guarantee a relatively small correlation time without losing much accuracy. Its optimal value depends on the potential energy surface of the system[1], a reasonable choice is $\tilde{\tau}_{\mathrm{NHC}} \geq 20\Delta t$.[2] $M_{NHC} = 4$ coupling thermostats are used in each chain of NHC in the simulations throughout the paper. We note that an early analysis on non-Hamiltonian dynamics[3] relates the compressibility to the convergence of phase space sampling.

While investigating the parameter(s) in a thermostat for a system, one often considers the characteristic time for the potential autocorrelation function [as defined in Appendix B in the paper]. The smaller the $\tau_{UU}$ is, the more efficient the algorithm is for sampling the coordinate space[4, 5]. Similarly, one can define the characteristic time of the Hamiltonian autocorrelation function $\tau_{HH}$, which is a reasonable quantity to estimate the efficiency for sampling the phase space[6].

For general systems, it is expected that the optimal collision frequency in the Andersen thermostat is about $\sqrt{2}$ times of the optimal friction coefficient in Langevin dynamics, as a generalization of the conclusion in Appendix B. The range for the collision frequency of the Andersen thermostat or that for the friction coefficient of Langevin dynamics is quite broad for achieving reasonable accuracy when the 'middle' scheme is employed. When the characteristic frequency of the system is $\bar{\omega}$, the range for either thermostat parameter is recommended to be about $0.1\bar{\omega} \sim 10\bar{\omega}$ while considering both the sampling efficiency and accuracy.

**Tables and Figures**

**Table S1.** Optimal parameters for minimal correlation time for potential energy.

| System | $\nu_{opt}$ Andersen | $\gamma_{opt}$ Langevin | $\tau_{opt}^{-1}$ NHC | $(20dt)^{-1}$ NHC | $\nu_{opt} / \gamma_{opt}$ |
|---|---|---|---|---|---|
| $U(x) = m\omega^2 x^2 / 2$ | 1.4 au | 1.0 au | 1.1 au | 0.5 au | 1.4 |
| $U(x) = x^4 / 4$ | 1.0 au | 0.7 au | 0.6 au | 0.5 au | 1.43 |
| $H_2O$ | 0.62 $fs^{-1}$ (0.016 au) | 0.45 $fs^{-1}$ (0.011 au) | 0.74 $fs^{-1}$ (0.018 au) | 0.21 $fs^{-1}$ (0.005 au) | 1.36 |
| $(Ne)_{13}$ | 1.8E-3 $fs^{-1}$ | 1.3E-3 $fs^{-1}$ | None | 0.5 $fs^{-1}$ | 1.38 |

**Table S2.** Recommended range for thermostat parameters for acceptable accuracy

| System | $\nu_{opt}$ Andersen | $\gamma_{opt}$ Langevin | $\tau_{opt}^{-1}$ NHC |
|---|---|---|---|
| $U(x)=m\omega^2x^2/2$ | Whatever | Whatever | Whatever |
| $U(x)=x^4/4$ [a] | ≥0.1 au | ≥0.01 au | 0.01-0.4 au |
| $H_2O$[b] | ≥0.04 $fs^{-1}$ | ≥0.04 $fs^{-1}$ | 0.004-0.4 $fs^{-1}$ |
| $(Ne)_{13}$[c] | ≥0.002 $fs^{-1}$ | ≥0.0006 $fs^{-1}$ | 0.0008-0.003 $fs^{-1}$ |

[a]The time interval $\Delta t = 0.5$ au.

[b]The time interval $\Delta t = 1.2$ fs.

[c]The time interval $\Delta t = 50$ fs.

**Figure Captions**

**Fig. S1** (Color). MD results using the 'middle' scheme for the correlation time of potential energy with respect to thermostat parameter ($\nu$ for Andersen thermostat, $\gamma$ for Langevin dynamics, and $\tau^{-1}$ for NHC thermostat). (a) The harmonic potential at $\beta = 8$. [Unit: atomic units (au)] (b) The quartic potential at $\beta = 8$. (Unit: au) (c) for $H_2O$ at T=100 K. (Unit: fs) (d) for $(Ne)_{13}$ at T=14 K. (Unit: fs). The unit of the thermostat parameter is au in Panels (a)-(b), while that is per femtosecond ($fs^{-1}$) in Panel (c)-(d). Statistical error bars are included.

**Fig. S2** (Color). The absolute difference from the exact/converged potential energy divided by the system temperature $\frac{|\Delta E|}{k_B T}$ for the quartic potential at $\beta = 8$. (a) Andersen thermostat (b) Langevin dynamics (c) NHC thermostat. The unit for thermostat parameters is atomic unit (au). Statistical error bars are included.

**Fig. S3** (Color). Same as Fig. S2 but for $H_2O$ at T=100 K. The unit for thermostat parameters is per femtosecond ($fs^{-1}$).

**Fig. S4** (Color). Same as Fig. S3 but for $(Ne)_{13}$ at T=14 K.

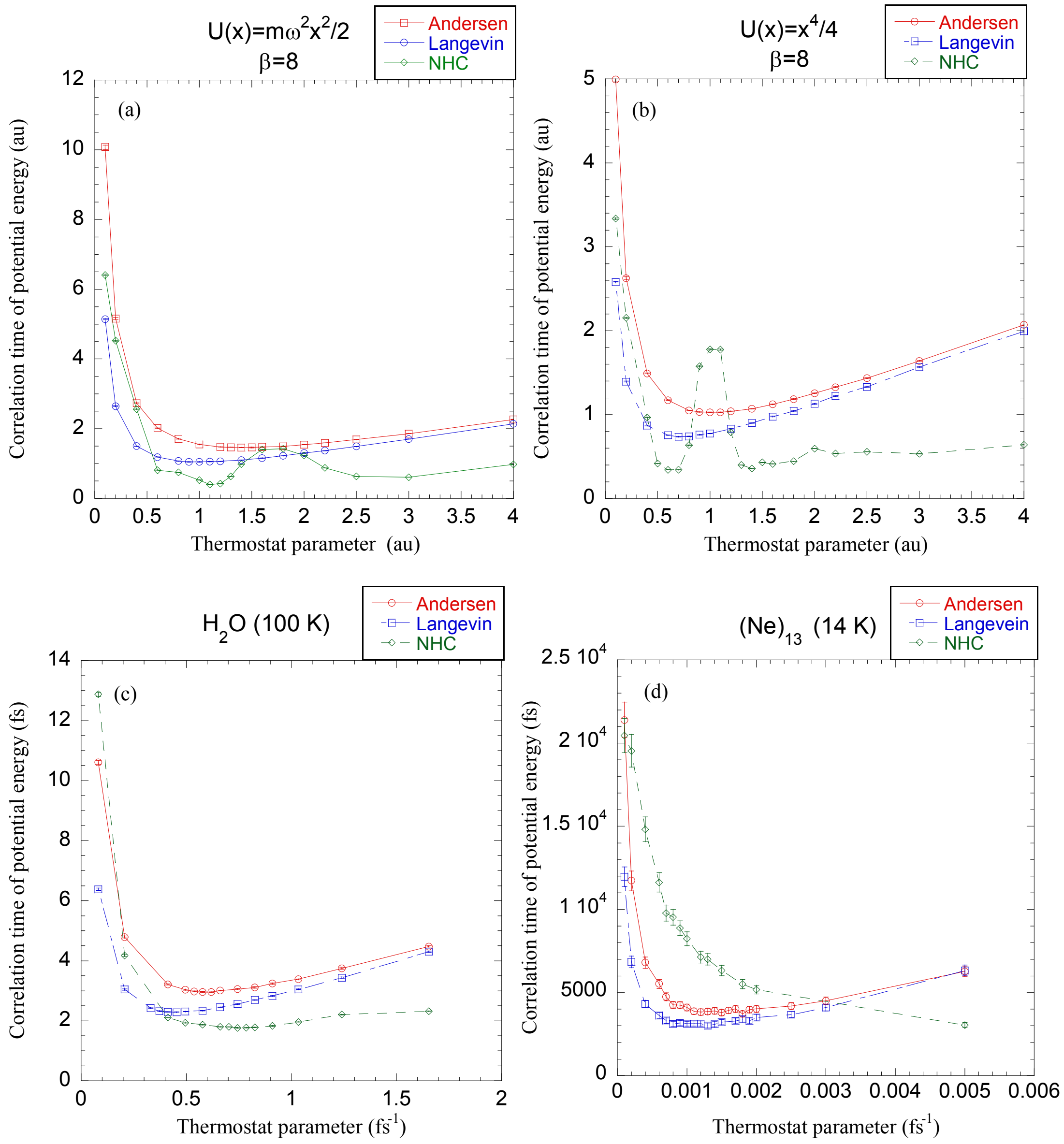

**Fig. S1**

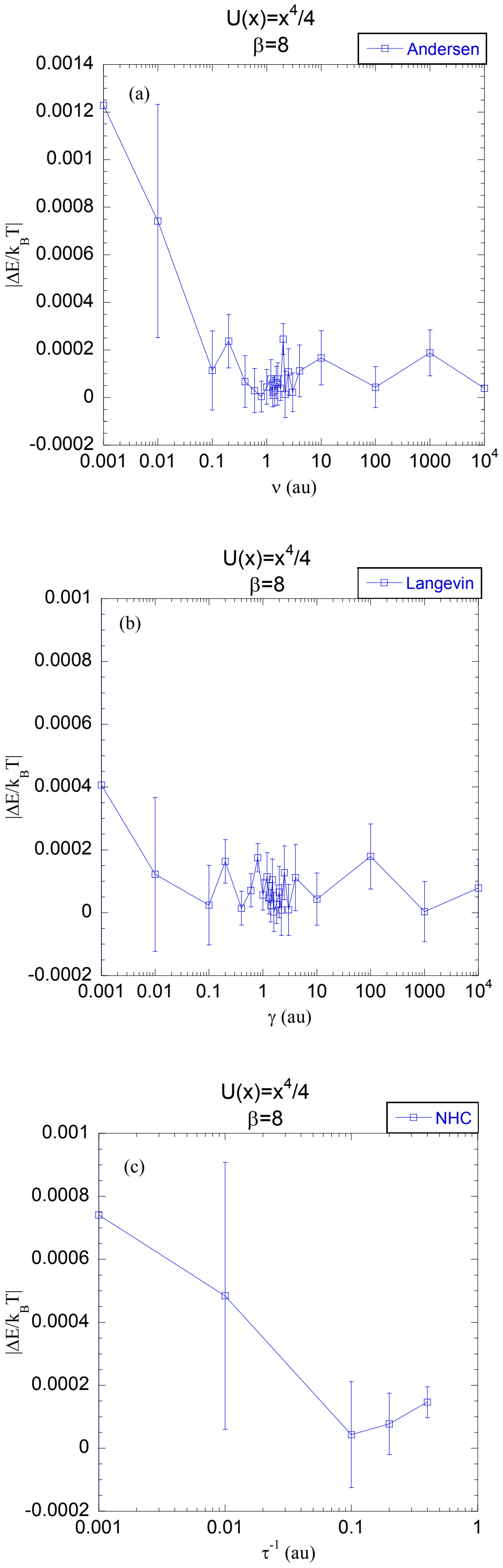

U(x)=x4/4
β=8
Andersen
(a)
|ΔE/kBT|
0.0014
0.0012
0.001
0.0008
0.0006
0.0004
0.0002
0
-0.0002
0.001 0.01 0.1 1 10 100 1000 104
ν (au)
U(x)=x4/4
β=8
Langevin
(b)
|ΔE/kBT|
0.001
0.0008
0.0006
0.0004
0.0002
0
-0.0002
0.001 0.01 0.1 1 10 100 1000 104
γ (au)
U(x)=x4/4
β=8
NHC
(c)
|ΔE/kBT|
0.001
0.0008
0.0006
0.0004
0.0002
0
-0.0002
0.001 0.01 0.1 1
τ-1 (au)

**Fig. S2**

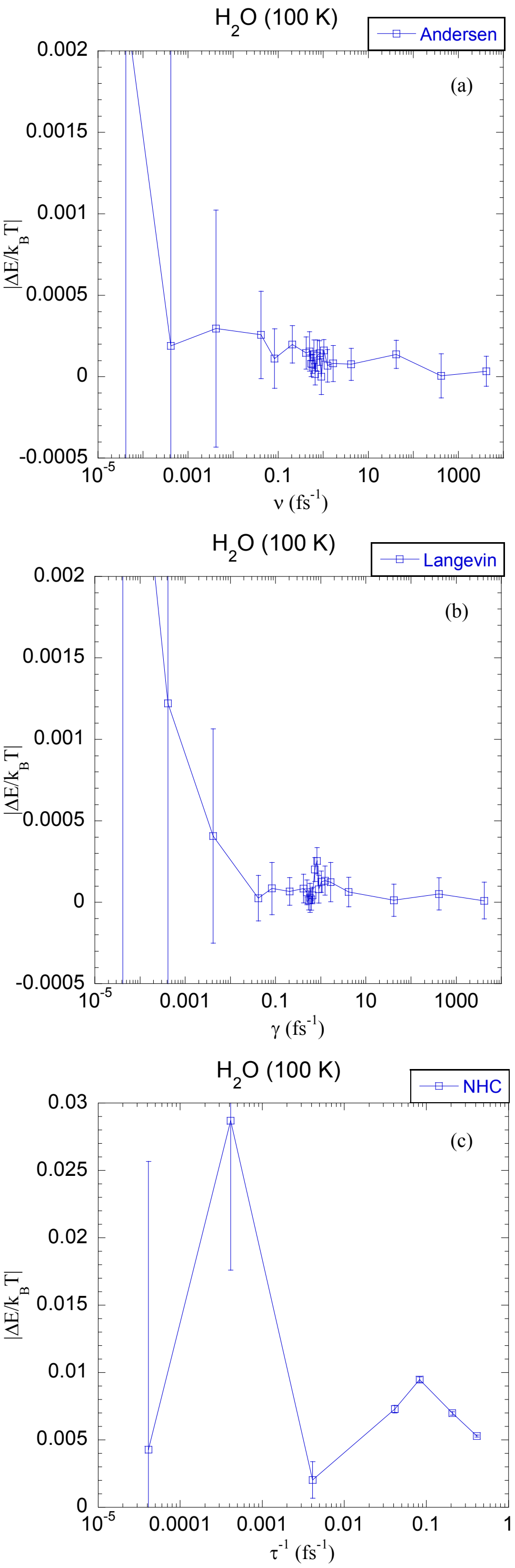

H2O (100 K)
Andersen
(a)
|ΔE/kBT|
0.002
0.0015
0.001
0.0005
0
-0.0005
10-5
0.001
0.1
10
1000
ν (fs-1)
H2O (100 K)
Langevin
(b)
|ΔE/kBT|
0.002
0.0015
0.001
0.0005
0
-0.0005
10-5
0.001
0.1
10
1000
γ (fs-1)
H2O (100 K)
NHC
(c)
|ΔE/kBT|
0.03
0.025
0.02
0.015
0.01
0.005
0
10-5
0.0001
0.001
0.01
0.1
1
τ-1 (fs-1)

**Fig. S3**

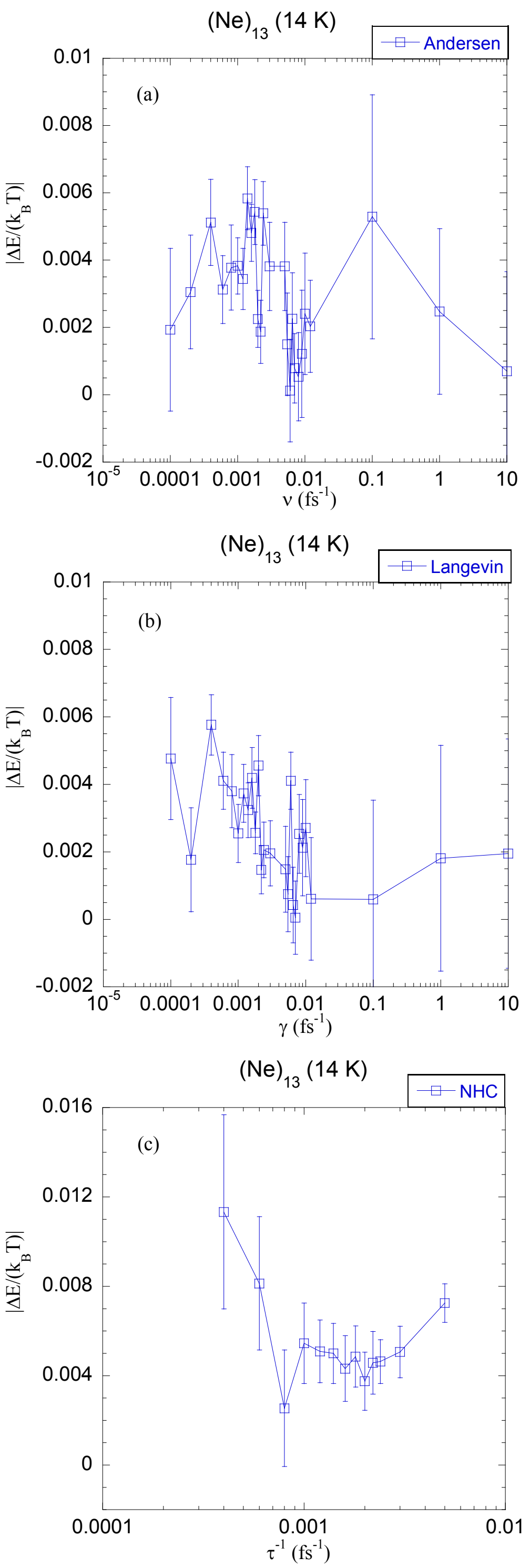
(Ne)13 (14 K)
Andersen
(a)
|ΔE/(kBT)|
0.01
0.008
0.006
0.004
0.002
0
-0.002
10-5
0.0001
0.001
0.01
0.1
1
10
ν (fs-1)
(Ne)13 (14 K)
Langevin
(b)
|ΔE/(kBT)|
0.01
0.008
0.006
0.004
0.002
0
-0.002
10-5
0.0001
0.001
0.01
0.1
1
10
γ (fs-1)
(Ne)13 (14 K)
NHC
(c)
|ΔE/(kBT)|
0.016
0.012
0.008
0.004
0
0.0001
0.001
0.01
τ-1 (fs-1)

**Fig. S4**

## ACKNOWLEDGMENT

This work was supported by the Ministry of Science and Technology of China (MOST) Grant No. 2016YFC0202803, by the National Science Foundation of China (NSFC) Grants No. 21373018 and No. 21573007, by the Recruitment Program of Global Experts, by Specialized Research Fund for the Doctoral Program of Higher Education No. 20130001110009, and by Special Program for Applied Research on Super Computation of the NSFC-Guangdong Joint Fund (the second phase). We acknowledge the Beijing and Tianjin supercomputer centers for providing computational resources.